\documentclass[namedreferences]{solarphysics}

\usepackage[hyperref,optionalrh,showbiblabels]{spr-sola-addons} 
\usepackage{graphicx}        
\usepackage{color}           
\usepackage{breakurl}        
\usepackage{placeins}

\renewcommand{\vec}[1]{{\mathbfit #1}}
\newcommand{\deriv}[2]{\frac{{\mathrm d} #1}{{\mathrm d} #2}}

\newcommand{\uvec}[1]{ \hat{\mathbf #1} }

\newcommand{\ap}{ \vec A_p}

\newcommand{\bb}{\vec B}

\chardef\us=`\_

\begin{document}

\begin{article}
\begin{opening}

\title{Probing the effect of cadence on the estimates of photospheric  energy and helicity injections in eruptive active region NOAA AR 11158}

\author[addressref=aff1,corref,email={
erkka.lumme@helsinki.fi}]{\inits{E.}\fnm{E.}~\lnm{Lumme}\orcid{0000-0003-2045-5320}}
\author[addressref={aff2,aff3}]{\fnm{M.~D.}~\lnm{Kazachenko}\orcid{0000-0001-8975-7605}}
\author[addressref=aff3]{\fnm{G.~H.}~\lnm{Fisher}\orcid{0000-0002-6912-5704}}
\author[addressref={aff4,aff3}]{\fnm{B.~T.}~\lnm{Welsch}\orcid{0000-0003-2244-641X}}
\author[addressref=aff1]{\fnm{J.}~\lnm{Pomoell}\orcid{0000-0003-1175-7124}}
\author[addressref=aff1]{\fnm{E.~K.~J.}~\lnm{Kilpua}\orcid{0000-0002-4489-8073}}

\address[id=aff1]{Department of Physics, University of Helsinki, P.O. Box 68, FI-00014 Helsinki, Finland}
\address[id=aff2]{Astrophysical and Planetary Sciences, University of Colorado, 2000 Colorado Avenue, Boulder, CO 80309, USA}
\address[id=aff3]{Space Sciences Laboratory, University of California, 7 Gauss Way, Berkeley, CA 94720-7450, USA}
\address[id=aff4]{Natural and Applied Sciences, University of Wisconsin, Green Bay, Green Bay, WI 54311, USA}

\runningauthor{Lumme et al.}
\runningtitle{Probing the effect of cadence}

\begin{abstract}
In this work we study how the input data cadence affects the photospheric energy and helicity injection estimates in eruptive NOAA active region 11158. We sample the novel 2.25-minute vector magnetogram and Dopplergram data from the \emph{Helioseismic and Magnetic Imager} (HMI) instrument onboard the \emph{Solar Dynamics Observatory} (SDO) spacecraft to create input datasets of variable cadences ranging from 2.25 minutes to 24 hours. We employ state-of-the-art data processing, velocity and electric field inversion methods for deriving estimates of the energy and helicity injections from these datasets. We find that the electric field inversion methods that reproduce the observed magnetic field evolution through the use of Faraday's law are more stable against variable cadence: the PDFI (PTD-Doppler-FLCT-Ideal) electric field inversion method produces consistent injection estimates for cadences from 2.25 minutes up to 2 hours, implying that the photospheric processes acting on time scales below 2 hours contribute little to the injections, or that they are below the sensitivity of the input data and the PDFI method. On other hand, the electric field estimate derived from the output of DAVE4VM (Differential Affine Velocity Estimator for Vector Magnetograms), which does not fulfil Faraday's law exactly, produces significant variations in the energy and helicity injection estimates in the 2.25-minute to 2-hour cadence range. We present also a third, novel DAVE4VM-based electric field estimate, which corrects the poor inductivity of the raw DAVE4VM estimate. This method is less sensitive to the changes of cadence, but still faces significant issues for the lowest of considered cadences ($\geq$2 hours). We find several potential problems in both PDFI- and DAVE4VM-based injection estimates and conclude that the quality of both should be surveyed further in controlled environments.
\end{abstract}
\end{opening}

\section{Introduction}
	\label{S-Introduction} 

Estimates for energy and helicity injections from the photosphere to the upper solar atmosphere in active regions are important for studying the dynamics of flux emergence \citep{Cheung2014,Liu2014}, flux cancellation \citep{Welsch2006,Yardley2018}, and the evolution of active regions \citep{Van2015,Cheung2016}. These estimates are also found to be particularly important for determining when and how
solar eruptions, such as flares and coronal mass ejections (CMEs), occur in active regions \citep{Cheung2012,Tziotziou2013,Kazachenko2015,Pariat2017,Pomoell2019}. Despite the fact that the coronal energy and helicity budgets may be estimated via coronal modeling (though with significant uncertainties, e.g. \citealp{DeRosa2015}), there are also less computationally intensive methods for estimating these quantities: so-called evolutionary estimates for the energy and helicity injections are acquired by integrating the photospheric Poynting and relative helicity fluxes in space and time \citep{Kazachenko2015}. These fluxes, in turn, are estimated using the photospheric electric and plasma velocity fields, which can be inverted from the remote sensing observations of the photospheric magnetic field and line-of-sight (LOS) plasma velocity. Due to the often used simplifying assumption of ideal Ohm's law,
\begin{equation}
\label{Eq-ideal_Ohm}
\vec{E} = -\vec{V} \times \vec{B}
\end{equation}
in the photosphere, the electric and plasma velocity fields are interchangeable in this context.

The accuracy of these estimates has progressed significantly over the past decade or so both due to improved remote sensing observations as well as developments in the inversion methods. Photospheric vector magnetic field estimates (vector magnetograms) and LOS plasma velocity estimates (Dopplergrams), based on spectropolarimetric observations of the Zeeman and Doppler effects, are currently provided by several magnetographs \citep[see][for a review]{Lagg2017}, including the \emph{Helioseismic and Magnetic Imager} (HMI) \citep{Scherrer2012,Schou2012} onboard \emph{Solar Dynamics Observatory} (SDO) \citep{Pesnell2012}. In turn, a wide collection of inversion methods have been developed for determining the photospheric plasma velocity and/or electric field components \citep[see][for further details]{Welsch2007,Schuck2008,Ravindra2008,Kazachenko2014,Tremblay2015,Lumme2017}, and these methods have been tested using both real and synthetic input data. 

The state-of-the-art inversion methods that are currently publicly available include the PDFI electric field inversion method \citep{Kazachenko2014,Fisher2019} and the DAVE4VM velocity inversion method \citep{Schuck2008}. These methods have shown their accuracy in estimating the electric/velocity fields and the related energy and helicity fluxes in a single test case, which we refer to as the ``ANMHD test''. In the ANMHD test, synthetic magnetic field and LOS plasma velocity estimates from an anelastic magnetohydrodynamic (ANMHD) simulation \citep{Abbett2004} of an emerging flux rope are used as input to the inversion, and the inverted velocity/electric fields are compared with the known fields in the simulation \citep{Welsch2007,Schuck2008,Kazachenko2014}. The fact that only a single simulation test has been used potentially limits the generality of the results: for example, the vertical emergence of flux is overemphasized in the ANMHD simulation, which causes the role of shearing motions in the in the energy and helicity injection to be understated \citep{Welsch2007,Kazachenko2014}. Furthermore, the synthetic data from the simulation is also smooth both in time and space, and does not necessarily represent well actual vector magnetogram and Dopplergram input, which often exhibit small-scale structures, noise and other artefacts.

Despite the limitations in the ANMHD-based validation both the PDFI and DAVE4VM inversion methods have been used successfully with observational input to estimate photospheric energy and helicity fluxes \citep[e.g.][]{Liu2012,Kazachenko2015,Liu2016,Lumme2017,Bi2018}, to study the properties of the photospheric plasma velocity field \citep[e.g.][]{Liu2016nature,Wang2017}, and to constrain the data-driven boundary conditions for coronal simulations \citep{Fisher2015,Pomoell2019}. However, the use of the methods with actual observations has been mostly limited to a single type of vector magnetogram and Dopplergram input from the SDO/HMI instrument, thus being fixed to certain cadence (12 minutes), resolution (0.5'' per pixel) and noise-characteristics ($\sigma_B \sim 100$ Mx cm$^{-2}$) of that data. Thus, it is unclear how the PDFI and DAVE4VM methods respond to, for example, different cadence, resolution and noise characteristics of the input data. It is also not known whether the possible undersampling of the photospheric evolution (i.e. insufficient temporal cadence for capturing the photospheric motions in a given spatial resolution) in the SDO/HMI data results in a loss of physical processes that have a significant contribution to the injection of energy and helicity. Moreover, \citet{Leake2017} found that undersampling may result in spurious energy fluxes into the corona.

Considering the crucial importance of energy and helicity injection estimates as well as the variety of dataset and products available now and in the future, an assessment of how the inversion methods respond to different input data is needed. In this paper we perform a comprehensive study on the response of energy and helicity injection estimates to input data of which the temporal resolution is varied. We sample the novel 135-second high-cadence vector magnetogram and Dopplergram data input from the SDO/HMI instrument \citep{Sun2017} as well as the nominal 12-minute data \citep{Hoeksema2014} to create an ensemble of input datasets with variable cadence ranging from 135 seconds all the way up to 24 hours, thus covering cadences of data products from other instruments \citep[see e.g.][for review]{Lagg2017}. We invert the photospheric velocity and electric fields from this data using the PDFI and DAVE4VM methods. We develop new self-consistent ways to optimize the inversion methods for each cadence and discuss also the role of spatial resolution and undersampling at the lowest cadences. 

Our study focuses on an eruptive active region NOAA AR 11158, thus making our results relevant for studies of active regions and solar eruptions. This active region was chosen because it is located close to the disk center all the way from its emergence to its strongest activity (X2.2 flare and a halo CME), which ensures good data quality and spatial resolution for studying the energy and helicity injection over this period. The region has also been extensively studied offering us a baseline of results for reference and context \citep[e.g.][]{Schrijver2011,Cheung2012,Liu2012,Sun2012,Tziotziou2013,
Kazachenko2015,Fisher2015,Lumme2017,Inoue2018}.

This paper is organized as follows: In Section 2 we present the used data products, inversion methods, and the approaches we use for optimizing the methods for use with input data of variable cadence. Section 3 details our central findings on the effect of cadence on the energy and helicity injection estimates. Section 4 discusses the observed limitations and issues in our results as well as the implications of our findings on the applicability of the inversion methods in estimating the photospheric energy and helicity injections. Section 5 summarizes our results and conclusions.

\section{Data and methods}
	\label{S-Data_and_methods} 

In this section we describe how we download and process the vector magnetogram and Dopplergram data to create data series of variable cadence and spatial resolution for NOAA AR 11158. We then discuss how we optimize and employ optical flow methods to produce additional estimates for the plasma velocity, and then invert the photospheric electric field from this data. Finally, we present our method for deriving the energy and helicity injections and their error estimates.

\subsection{Processing of vector magnetogram and Dopplergram data}
	\label{S-proc_magnetograms}
	
As the input data for this work we use full-disk disambiguated vector magnetograms and Dopplergrams from the SDO/HMI instrument \citep{Schou2012,Scherrer2012}, which we download from the Joint Science Operations Center (JSOC, \url{http://jsoc.stanford.edu/}). We use both the nominal 12-minute (720 s) data \citep{Hoeksema2014} as well as the novel 2.25-minute (135 s) data \citep{Sun2017}. Both datasets have the same spatial resolution (0.5'' per pixel in the plane of sky), but differ in the methods used to process Stokes vector data for the magnetic field inversion, the disambiguation of the azimuth, and in the noise levels (see \citealp{Sun2017}, for details). Vector magnetograms in these datasets are disambiguated in the strong field pixels (thresholds $|\vec{B}| \sim 200$ Mx cm$^{-2}$ and $|\vec{B}| \sim 150$ Mx cm$^{-2}$ for 2.25- and 12-minute data, respectively) using the minimum energy method \citep{Metcalf1994}. In the weak field pixels three less sophisticated methods are offered as user-defined options \citep{Hoeksema2014}, from which we choose to use the random disambiguation (as recommended e.g. by \citealp{Liu2017}; see also discussion in \citealp{Lumme2017}). 

For Dopplergrams (i.e. LOS plasma velocity maps) we use the $V_{Inv}$ velocity provided by the magnetic field inversion of the vector magnetogram datasets above. This differs from the work of \citet{Kazachenko2015}, who employed the $V_{Dop}$ estimate based on the simpler ``MDI-like algorithm'' (see \citealp{Hoeksema2014} for details about the differences between $V_{Inv}$ and $V_{Dop}$). Similarly to \citet{Kazachenko2015} we calibrate the Dopplergrams using the magnetic calibration method of \citet{Welsch2013}, which removes the observer motion, solar rotation and convective blueshift from the data; the convective blueshift bias velocity is determined as the median of Dopplergram velocities over all pixels in polarity inversion lines (PILs) sufficiently close to the disk center ($<60^{\circ}$ in heliocentric angle). Before subtracting the constant blueshift bias velocity from full-disk Dopplergrams we smooth these velocities in time using a temporal smoother with a width of 4.2 hours \citep{Kazachenko2015} to reduce temporal noise in the bias velocities.

\begin{figure}[htb]   
    \centerline{\includegraphics[width=\textwidth,trim = 0.0cm 1.0cm 0cm 1.0cm]{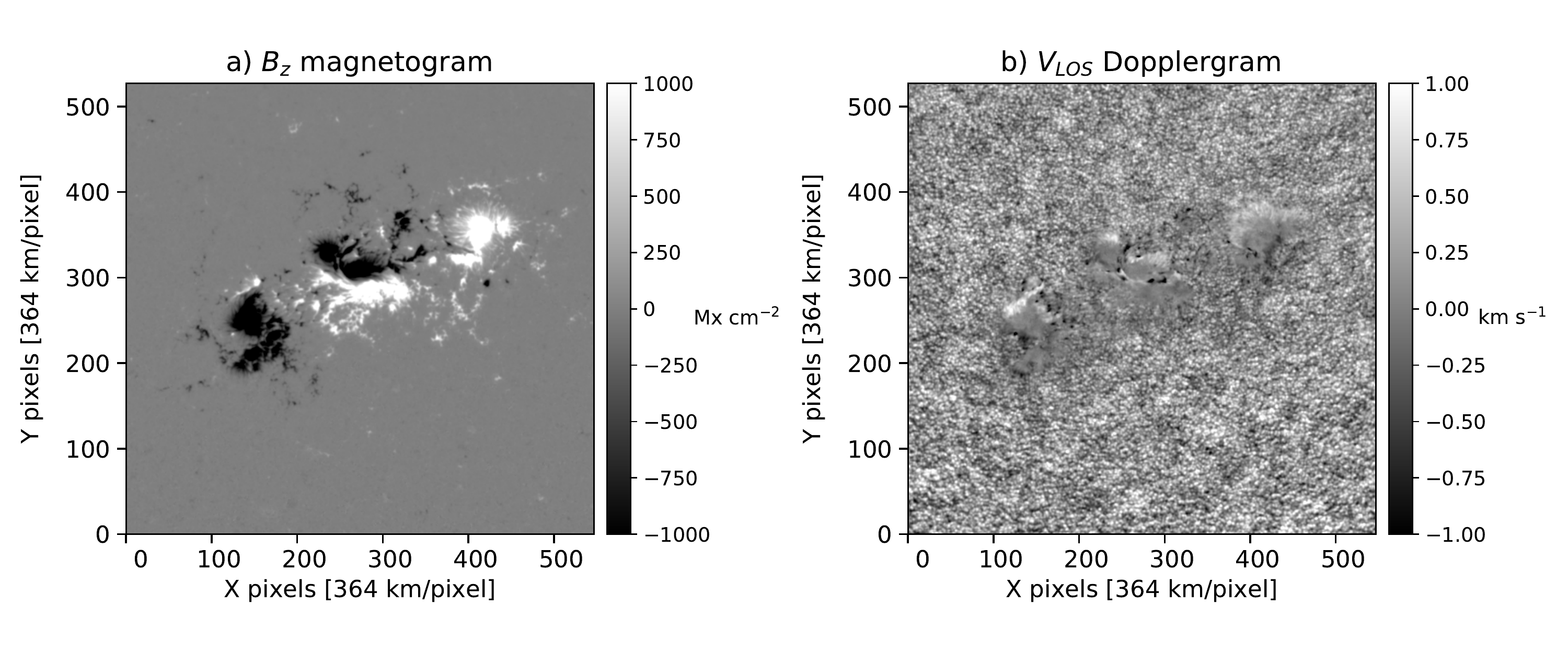}}  
   \caption{Example $B_z$ magnetogram and $V_{LOS}$ Dopplergram taken from our high-cadence 2.25-minute reprojected series that tracks the NOAA active region 11158. Note that contrary to the usual convention $V_{LOS}$ sign is negative for downflows of plasma. The frames show the active region on Feb 14 01:48 UT, 2011 close to the central meridian passage of the active region, and magnetogram and Dopplergram data near this time was used as a representative input for optimization of the velocity inversion parameters (Section \ref{S-vel_inversion}) as well as for deriving the error estimates for energy and helicity injections (Section \ref{S-error_analysis}).} 
   \label{F-Bz_examp_frame}
\end{figure}

After downloading and processing the full-disk vector magnetograms and Dopplergrams we reproject the magnetograms to a local Cartesian frame and vector basis $(B_x,B_y,B_z)$ using Mercator projection that tracks the NOAA active region 11158 over its disk transit (see Figure \ref{F-Bz_examp_frame} for example frames from this time series). We use here methods described in \citet{Lumme2017}, and this processing includes also removal of bad pixels of the magnetic field inversion \citep{Hoeksema2014} and spurious temporal flips in the azimuth disambiguation \citep{Welsch2013}, where we use identical parameters for both 2.25- and 12-minute data (see \citealp{Lumme2017} for details). Dopplergram data is interpolated to the same system as the magnetograms, keeping the information about the LOS direction for each pixel. The resulting active region patch has $547 \times 527$ pixels (with a projection pixel size of 0.03$^{\circ}$, 0.5'' at the disk center, $\sim$364 km on the Sun), and the series of reprojected magnetograms spans from Feb 10 14:00 UT to Feb 17 00:00 UT.

\subsection{Sampling the vector magnetogram and Dopplergram data}
	\label{S-sampling_data}
	
A direct comparison on the effect of cadence between the 2.25- and 12-minute vector magnetogram and Dopplergram datasets is problematic, as they have differences in the Stokes vector data processing and in the noise levels of the magnetic field \citep{Sun2017}. In order to mitigate the effect of these differences we created 6 datasets with variable cadence by sampling the 2.25-minute data only. However, the 12-minute data was still included in the study as a reference due to its general use in previous works and its better availability. The cadences created by sampling the 2.25-minute data are: 2.25 minutes (1$\times$2.25-min), 11.25 minutes (5$\times$2.25-min), 2.025 hours (54$\times$2.25-min), 6 hours (160$\times$2.25-min), 12 hours (320$\times$2.25-min) and 24 hours (640$\times$2.25-min), where the cadences $>$2.25-min are created from the 2.25-minute data by taking every 5th, 54th, 160th,... frame from the 2.25-minute series. Hereafter we refer to the different cases by their cadence (e.g. ``6 hours''), except for the 11.25-min case, which we refer also to as ``mock nominal'' cadence, as it mimics the nominal 12-minute HMI cadence, but with noise characteristics and Stokes data processing consistent with the 2.25-min magnetogram and Dopplergram data.

It is evident that the lowest cadences ($\geq$2 h) undersample the data significantly when considering the typical photospheric motions of $V_h \lesssim 1$ km s$^{-1}$, for which we get: $V_h\Delta t/\Delta x \gtrsim 20$, for cadences $\Delta t \geq 2$ h and $\Delta x = 364$ km (pixel size in our vector magnetogram series). This undersampling is expected to introduce issues particularly for the optical flow velocity methods (Section \ref{S-vel_inversion}) and may even introduce spurious energy and helicity fluxes \citep[see][]{Leake2017}. In order to study this effect in our results we perform additional analysis for the lowest cadences ($\geq$2 h), for which we create new data series by rebinning the data by a factor of 15 to yield a projected pixel size of $0.45^{\circ}$ that corresponds $\Delta x' \approx 5470$ km and $V_h\Delta t/\Delta x \gtrsim 1.3$. (The amount of undersampling for the rebinned data is discussed further in Appendix \ref{S-opt_res_FLCT_and_D4VM}.)

Finally, for the optical flow velocity inversion (Section \ref{S-vel_inversion}), the electric field inversion (Section \ref{S-electric_f_inv}), and for computing the energy and helicity fluxes (Section \ref{S-est_en_hel_inj}) we mask the noise-dominated pixels from the magnetograms to avoid spurious effects caused by noise \citep[see][for further discussion]{Kazachenko2015,Lumme2017}. For this purpose we use a constant noise threshold of $|\vec{B}| = 300$ Mx cm$^{-2}$ for all datasets discussed above. This threshold was used previously for 2.25-minute HMI data by \citet{Sun2017}. 

\subsection{Optical flow velocity inversion}
	\label{S-vel_inversion}
	
Having determined all components of the photospheric magnetic field and one (LOS) component of plasma velocity from the processed data products (Sections \ref{S-proc_magnetograms} and \ref{S-sampling_data}) we need two more velocity components to fully constrain the electric field from the ideal Ohm's law (Eq. \ref{Eq-ideal_Ohm}). We estimate (a part of) the missing velocity components using optical flow methods, which track motion of features in the magnetograms. In this study we use two optical flow methods, each linked to a specific electric field inversion method. The first method is the Fourier Local Correlation Tracking (FLCT) method \citep{Fisher2008} that estimates the horizontal velocity components parallel to the photosphere $\vec{V}_h = (V_x,V_y)$, which are then used as a part of the PDFI electric field inversion method (Section \ref{S_einv_PDFI}). The second method is the Differential Affine Velocity Estimator for Vector Magnetograms (DAVE4VM) \citep{Schuck2008} that provides the full three-component velocity field $\vec{V} = (V_x,V_y,V_z)$, which is then used to derive the two DAVE4VM-based electric field estimates (Section \ref{S_einv_DAVE4VM}). 

The FLCT method determines the horizontal optical flow $\vec{V}_h$ between two $B_z$ magnetograms ($B_z(x_i,y_j,t_1)$ and $B_z(x_i,y_j,t_2)$) at each pixel $(x_i,y_j)$ by finding a shift $(\delta x,\delta y)$ that maximizes the cross-correlation function $C^{i,j}(\delta x, \delta y)$ between the first image $S_1(x,y)$ and second image $S_2(x,y)$ -- both windowed to include only the neighborhood of the pixel in question:
\begin{eqnarray}
\label{Eq-FLCT_C_max}
C^{i,j}(\delta x,\delta y) & = \int \int dx \ dy \ S_1^{i,j*}(-x,-y) S_2^{i,j}(\delta x-x,\delta y-y), 
\end{eqnarray}
where 
\begin{eqnarray}
\label{Eq-FLCT_subimage}
S_k^{i,j}(x,y) = B_z(x,y,t_k) \exp\left[-\frac{(x-x_i)^2+(y-y_j)^2}{\sigma_{FLCT}^2}\right]
\end{eqnarray}
is the windowed subimage centered at the pixel $(x_i,y_j)$; the width of the Gaussian windowing function $\sigma_{FLCT}$ is a free parameter of the method. The optimal shift $(\delta x,\delta y)$ is determined to subpixel accuracy \citep[see][for details]{Fisher2008}, and it gives the velocity:
\begin{eqnarray}
\label{Eq-V_h_FLCT}
\vec{V}_h^{i,j}((t_1+t_2)/2) & = \frac{1}{\Delta t}(\delta x,\delta y)\bigg|_{\lbrace(\delta x, \delta y) | \max [C^{i,j}(\delta x,\delta y)]\rbrace},
\end{eqnarray}
where $\Delta t = t_2 - t_1$ is the temporal distance between the magnetograms.

As discussed by \citet{Schuck2005} the velocity of Eq. \ref{Eq-V_h_FLCT} determined by maximizing the correlation in Eq. \ref{Eq-FLCT_C_max} fulfills (in a least squares sense) the advection equation:
\begin{eqnarray}
\label{Eq-adv_eq_FLCT}
\frac{\partial B_z}{\partial t} + \vec{V}_h \cdot \nabla_h B_z = 0,
\end{eqnarray}
weighted by the Gaussian windowing function of Eq. \ref{Eq-FLCT_subimage}. In the DAVE4VM method this minimization approach is generalized for the normal ($z$) component of Faraday's law under the assumption of the ideal Ohm's law:
\begin{eqnarray}
\label{Eq-n_induc_eq}
\frac{\partial B_z}{\partial t} = -[\nabla \times \vec{E}]_z = [\nabla \times (\vec{V} \times \vec{B})]_z = -\nabla_h \cdot (B_z\vec{V}_h - V_z \vec{B}_h),  
\end{eqnarray}
and the method determines a full three-component velocity field $\vec{V} = (V_x,V_y,V_z)$ at each pixel $(x_i,y_i)$ that minimizes the deviations to the normal component of the induction equation (Eq. \ref{Eq-n_induc_eq}) in a least squares sense \citep{Schuck2008}. The input for this minimization consists of the time derivative of the $B_z$ component $\partial B_z/\partial t$ and the spatial derivatives of the magnetic field ($\partial_l B_m$, $l \in \{x,y\}$ and $m \in \{x,y,z\}$) determined from a time series of vector magnetograms. The input data are windowed similarly to Eq. \ref{Eq-FLCT_subimage}, but instead of a Gaussian windowing function, DAVE4VM employs a square top hat as the windowing function. The side length of the top hat is a free parameter, which we hereafter refer to as the ``DAVE4VM window size''. 

When specifying the input data for FLCT and DAVE4VM, we employ a central difference scheme where magnetograms at times $t$ and $t \pm \Delta t$ are used to estimate the velocity at time $t$ (see Appendix \ref{S-time_dscr_v_and_einv}). We also remove the noise-dominated pixels ($|\vec{B}| < 300$ Mx cm$^{-2}$) from the output (see Appendix \ref{S-mask_in_v_and_einv}).

Similarly to the work of \citet{Schuck2008} and \citet{Liu2012} we choose the DAVE4VM window size so that the output velocity $\vec{V}$ optimally reproduces the normal component of the induction equation (Eq. \ref{Eq-n_induc_eq}) over the entire magnetogram, excluding the noise-dominated pixels $|\vec{B}| < 300$ Mx cm$^{-2}$. The optimal $\sigma_{FLCT}$ parameter is chosen in a similar fashion so that the output velocity best optimizes the advection equation for the $B_z$ component of the magnetic field (Eq. \ref{Eq-adv_eq_FLCT}) \citep[as suggested by][]{Fisher2008}, again excluding the masked noise-dominated pixels ($|B_z| \lesssim 300$ Mx cm$^{-2}$). Further details of the optimization and the analysis of the results are presented in Appendix \ref{S-opt_res_FLCT_and_D4VM}. The optimization results for all of our input datasets are collected in Table \ref{T-optimization_results}.

After specifying the optimal FLCT and DAVE4VM parameters for all of our vector magnetogram datasets (see Section \ref{S-ind_DAVE4VM_E_and_H_inj}), we then run the velocity inversion codes for all input magnetogram time series of various cadence and spatial resolution (Section \ref{S-sampling_data}) producing $\vec{V}_h$(FLCT) and $\vec{V}$(DAVE4VM) time series for each to be used in the electric field inversion.

\subsection{Electric field inversion}
	\label{S-electric_f_inv}
	
After producing time series of the photospheric vector magnetogram, Dopplergram and optical flow velocity estimates for variable cadence and spatial resolution, we compute the photospheric electric field using each of the input data series and three inversion schemes: one based on the direct use of the PDFI method and two based on the DAVE4VM velocity inversion. These are presented in detail below. The temporal and spatial discretization used in the inversion methods below as well as the masking of the noise-dominated pixels ($|\vec{B}| < 300$ Mx cm$^{-2}$) are detailed in Appendix \ref{S-tech_detail_appendix}.

\subsubsection{PDFI method}
	\label{S_einv_PDFI}

The PTD-Doppler-FLCT-Ideal (PDFI) method \citep{Kazachenko2014,Fisher2012} is a comprehensive, publicly available approach for photospheric electric field inversion, and the method employs all established types of input data:  magnetograms, Dopplergrams and optical flow estimates. The approach is based on the decomposition of the electric field into the inductive $\vec{E}_I$ and non-inductive component $-\nabla \psi$:
\begin{equation}
\label{Eq-ind_nind_decomp}
\vec{E} = \vec{E}_I - \nabla \psi.
\end{equation}
The inductive (divergence-free) component is constrained by Faraday's law:
\begin{equation}
\label{Eq-Far_law}
\nabla \times \vec{E} = \nabla \times \vec{E}_I = -\frac{\partial \vec{B}}{\partial t},
\end{equation}
where $\partial \vec{B}/\partial t$ may be estimated from a time series of vector magnetograms. In the PDFI method the inductive component is solved using the Poloidal-Toroidal Decomposition (PTD) \citep{Fisher2010,Chandrasekhar1970}, and the resulting electric field fulfills the $z$ component of the Faraday's law exactly -- except for small numerical errors detailed in Appendix \ref{S-app_mods_to_PDFI}. The non-inductive (curl-free) component requires additional constraints from the data that may be retrieved from the ideal Ohm's law and the velocity estimates:
\begin{eqnarray}
\label{Eq-div_E_Ohm}
\nabla^2 \psi = -\nabla \cdot \vec{E} = \nabla \cdot (\vec{V} \times \vec{B}) 
\end{eqnarray}
In the PDFI method the Poisson equation above is not solved as such, but is instead split into three components: the Doppler contribution (D), FLCT contribution (F) and Ideal (I) contribution. The Doppler and FLCT contributions are deduced from Dopplergram and FLCT optical flow velocity estimates with spatial weighting, whereas the ideal contribution ensures that the total electric field (Eq. \ref{Eq-ind_nind_decomp}) is perpendicular to the magnetic field as implied by the ideal Ohm's law \citep[see][for details]{Kazachenko2014,Fisher2019}.

We employ the latest version of the PDFI method, the PDFI\_SS software (\url{http://cgem.ssl.berkeley.edu/cgi-bin/cgem/PDFI_SS/index}), which has several changes as compared to the original method described in \citet{Kazachenko2014}. The updates include the use of a staggered grids and spherical coordinates (``SS'' suffix stands for ``spherical staggered''), and they are presented in detail by \citet{Fisher2019} and summarized in Appendix \ref{S-app_mods_to_PDFI}. 

Since our input data patch is small (latitudinal half-width of the region $\sim 8^{\circ}$), distortion effects caused by the use of Mercator projection remain small: $1-\cos^2 8^{\circ} \approx 2$\% \citep[see e.g.][for details]{Kazachenko2015}. Therefore we get little benefit from using spherical coordinates in our analysis. Moreover, the DAVE4VM velocity inversion (Section \ref{S-vel_inversion}) operates on a Cartesian plane by default, and transforming into spherical coordinates would require modifications to the procedure. Therefore we decide to remain in the Cartesian approximation, which however, requires modifications to the use of the PDFI\_SS software that employs spherical coordinates by default. These modifications are detailed in Appendix \ref{S-app_mods_to_PDFI}.

\subsubsection{DAVE4VM-based methods}
	\label{S_einv_DAVE4VM}

Since DAVE4VM provides all three components of the velocity fields for the given vector magnetogram input, the output can be used to estimate the photospheric electric field directly from the ideal Ohm's law (Eq. \ref{Eq-ideal_Ohm}), $\vec{E} = -\vec{V}_D \times \vec{B}$, where $\vec{V}_D$ is the DAVE4VM velocity estimate. We use this as our first DAVE4VM-based electric field estimate and hereafter refer to it as \textbf{\emph{``raw DAVE4VM estimate''}} or \textbf{\emph{``raw DAVE4VM electric field''}}. 

The raw DAVE4VM estimate is the easiest to acquire, but as already noted by \citet{Schuck2008} it is not necessarily inductive, i.e. the raw DAVE4VM electric field does not necessarily fulfill the normal component of Faraday's law (or the normal component of the ideal induction equation). This inconsistency arises from two facts: First, the DAVE4VM method was intentionally formulated so that the output velocity fulfills the normal component of the induction equation (Eq. \ref{Eq-n_induc_eq}) only ``statistically within the window by minimizing the mean squared deviations from the ideal induction equation'' \citep{Schuck2008}. This is based on the idea that real magnetograms contain noise, and therefore complete inductiveness and thus also complete reproduction of the noise is not desired. Second, the minimization of the deviations to the induction equation is done at each pixel over a top-hat windowed subimage surrounding the pixel, and therefore the minimization problem at each pixel is different from the others. Thus, there is no guarantee or constraint that would force the velocity fields at neighboring pixels to yield inductivity in the chosen discretization of the induction equation. 

Although the DAVE4VM estimate is inductive to high accuracy in ANMHD tests \citep {Schuck2008}, as pointed out by \citet{Lumme2017} the inductivity of the raw DAVE4VM estimate is, however, very poor for real magnetogram input. As indicated by the results presented in Appendix \ref{S-opt_res_FLCT_and_D4VM}, a likely explanation for this is the larger spatial and temporal noise and higher spatial structuring in real magnetograms as compared to the very smooth ANMHD data. The fact that the field does not drive the magnetic field evolution as in the observations presents significant problems both for the physical interpretation of the field and for its potential use as a data-driven boundary condition for coronal simulations. To remedy this, we recomputed a second electric field estimate from the DAVE4VM output, in which the inductivity is ensured (similarly as suggested in Section 3.2 of \citealp{Schuck2008}). We employ here the decomposition of Eq. \ref{Eq-ind_nind_decomp}, from which the inductive component is solved using the machinery of the PDFI method, being thus equivalent to the inductive component $\mathbf{E}_I$ of our PDFI estimates (Section \ref{S_einv_PDFI}, Eqs. \ref{Eq-ind_nind_decomp} and \ref{Eq-Far_law}), whereas the horizontal components of the non-inductive component $-\nabla_h \psi$ are solved from:
\begin{eqnarray}
\label{Eq-pois_psi_DAVE4VM}
\nabla_h^2 \psi &=& -\nabla_h \cdot \vec{E} = \nabla_h \cdot (\vec{V}_D \times \vec{B})_h
\end{eqnarray}
where $\vec{E}$ is the raw DAVE4VM electric field and $\vec{V}_D$ is the DAVE4VM velocity. The Poisson equation is solved using the same numerical tools as in the PDFI method \citep{Kazachenko2014,Fisher2019}. We refer to the total electric field estimate,
\begin{equation}
\label{Eq-ind_DAVE4VM_e_field}
\vec{E} = \vec{E}_I - \nabla_h \psi 
\end{equation}
as the \textbf{\emph{``inductive DAVE4VM estimate''}} or \textbf{\emph{``inductive DAVE4VM electric field''}}. Note that $\partial \psi/\partial z = 0$ for the non-inductive component $-\nabla \psi = -\nabla_h \psi$ of the inductive DAVE4VM estimate.

\subsection{Estimates of the energy and relative helicity injections}
	\label{S-est_en_hel_inj}
	
Total injections of magnetic energy and relative helicity can be estimated by integrating the photospheric vertical Poynting and relative helicity fluxes both in space and time \citep{Berger1984,Demoulin2007,Liu2012,Kazachenko2015,Lumme2017}:
\begin{eqnarray} 
\label{Eq-E_injection} 
E_m(t)&=&\int_0^t dt' \ \deriv{E_m}{t'} = \int_0^t dt' \int \ dA \ S_z \nonumber \\
&=& \frac{1}{\mu_0} \int_0^t dt' \int \ dA \ (\vec{E} \times \bb) \cdot \uvec{z},  \\
\label{Eq-H_injection} 
H_R(t) &=& \int_0^t dt' \ \deriv{H_R}{t'} = -2 \int_0^t dt' \int \ dA \ (\ap \times \vec{E}) \cdot \uvec{z}.	
\end{eqnarray}

We use here time discretization consistent with the electric field inversion (see Appendix \ref{S-time_dscr_v_and_einv}). The vector potential $\ap$ is solved from the $B_z$ component using the same Poloidal-Toroidal Decomposition method as in solving the inductive electric field $\vec{E}_I$ used in the PDFI and inductive DAVE4VM electric field estimates (Section \ref{S-electric_f_inv} and Appendix \ref{S-app_mods_to_PDFI}, Eqs. \ref{Eq-ind_nind_decomp}, \ref{Eq-Far_law}, \ref{Eq-pois_psi_DAVE4VM}, \ref{Eq-ind_DAVE4VM_e_field}). The same $\ap$ is used to derive the helicity flux for all of our three electric field estimates. It is also worth noting that we do not include the zero-padding regions, which are added to the input data maps in the inversion, to the area integrals above \citep[the padding is added only for numerical convenience in order to acquire more stable results for the solutions of Poisson equations, see][]{Kazachenko2014}. 

Note that in computing the Poynting flux we must mask the input magnetogram data consistently with the electric field inversion (Section \ref{S-mask_in_v_and_einv}). However, when computing the helicity flux, the masking is already included in the computation of $\ap$ and $\vec{E}$.

\subsection{Error analysis}
	\label{S-error_analysis}
 
We estimate the error in our final total energy and helicity injection estimates arising from the magnetogram noise following the Monte Carlo approaches of \citet{Liu2012} and \citet{Kazachenko2015}. We select a representative frame from our NOAA AR 11158 time series, more specifically the frame on Feb 14 01:48 UT close to the central meridian passage of the active region, and perform 200 \citep{Liu2012} Monte Carlo realizations for the velocity and electric field inversion results for this frame and for each cadence; the first realization in the ensemble is the original unperturbed case. For each of the 199 additional realizations we perturb the input magnetogram data by adding random Gaussian noise to each pixel. For the data series created from the 12-minute HMI input data we perturb the magnetic field components $(B_x,B_y,B_z)$ with Gaussian noise of width $(\sigma_x,\sigma_y,\sigma_z) = (100,100,30)$ Mx cm$^{-2}$ using the values of \citet{Kazachenko2015} derived for SDO/HMI data in the same active region. Although we have determined the $(\sigma_x,\sigma_y,\sigma_z)$ values also for our own data series by fitting a Gaussian to the weak field core \citep{Kazachenko2015,Welsch2013,DeForest2007}, we find that the values are too strongly dependent on the choice of the weak field disambiguation method (Section \ref{S-proc_magnetograms}). Therefore we use the values of \citet{Kazachenko2015}, which are derived from a dataset where a consistent disambiguation method is used for all pixels. We employ larger perturbations for all data series created from the 2.25-minute HMI magnetogram input, and add $50/\sqrt{3}$ Mx cm$^{-2}$ to each of the $(\sigma_x,\sigma_y,\sigma_z)$ values of the 12-minute data, consistent with the 50 Mx cm$^{-2}$ larger noise level of $|\vec{B}|$ for 2.25-minute magnetograms \citep{Sun2017}. 

Using the perturbed data from the Monte Carlo runs, we estimate the errors for the energy $dE_m/dt$ and helicity $dH_R/dt$ injection rates (Eqs. \ref{Eq-E_injection} and \ref{Eq-H_injection}) by computing the standard deviation ($\sigma_{dQ/dt}, \ Q \in \{E_m, H_R\}$) over the 200-element ensemble. We determine a final relative error $\bar{\sigma}_{dQ/dt}$ estimate by computing the ratio: 
\begin{equation}
\label{Eq-rel_sigma_error}
\bar{\sigma}_{dQ/dt} = \frac{\sigma_{dQ/dt}}{\mu_{dQ/dt}}
\end{equation}
where $\mu$ is the mean over the ensemble.  

The errors of the energy and helicity injection rates ($\sigma_{dQ/dt} = \bar{\sigma}_{dQ/dt}dQ/dt$) are propagated in the time integration to yield the errors of the total injections $\sigma_{Q}(t)$ at each time $t$. This is done using the following scheme: First, the relative error estimates $\bar{\sigma}_{dQ/dt}$ above are interpreted as fixed constants over the entire time series. Second, all $dQ/dt(t_i)$ estimates in the time series are assumed to be independent from each other, i.e. covariances $\textrm{cov}(dQ/dt(t_i), dQ/dt(t_j))$ vanish for all pairs $t_i, t_j$. Using the basic properties of the variance and standard deviation \citep[e.g.][]{Christensen1996} and the fact that we use the trapedzoidal rule in the time integration of Eqs. \ref{Eq-E_injection} and \ref{Eq-H_injection}, 
\begin{equation}
Q(t_N) = \left\lbrace\frac{1}{2}\left[\deriv{Q}{t}(t_1) + \deriv{Q}{t}(t_N)\right] + \sum_{i=2}^{N-1} \deriv{Q}{t}(t_i)\right\rbrace \Delta t,
\end{equation}
we get:
\begin{eqnarray}
\sigma_{Q}(t_N) &=& \sqrt{\left\lbrace\frac{1}{4}\left[\sigma_{dQ/dt}^2(t_1) + \sigma_{dQ/dt}^2(t_N)\right] + \sum_{i=2}^{N-1} \sigma_{dQ/dt}^2(t_i)\right\rbrace \Delta t^2} \nonumber \\
\label{Eq-error_prop_in_tint}
 &=& \bar{\sigma}_{dQ/dt} \Delta t \sqrt{\frac{1}{4}\left[\deriv{Q}{t}(t_1)^2 + \deriv{Q}{t}(t_N)^2\right] + \sum_{i=2}^{N-1}\deriv{Q}{t}(t_i)^2}
\end{eqnarray}
where $\sigma_{dQ/dt} = \bar{\sigma}_{dQ/dt}dQ/dt$, $\bar{\sigma}_{dQ/dt}$ is our relative error estimate for the injection rate, and $\Delta t = (t_N - t_1)/(N-1)$ is the cadence of the data series.

We have collected the relative error estimates for the rate of changes ($dE_m/dt$, $dH_R/dt$) over all of our input datasets in Table \ref{T-error_results}. Further validation of the results and discussion can be found from Section \ref{S-error_bars_disc}.
\begin{table}[hbt]
\caption{Relative errors $\bar{\sigma}_{dE_m/dt}$, $\bar{\sigma}_{dH_R/dt}$ for the energy and helicity injection rates arising from the magnetogram noise. Values are derived for a representative frame on Feb 14 01:48 UT close to the central meridian passage of the active region using each of our magnetogram/Dopplergram datasets, and velocity and electric field inversion methods. 
}
\label{T-error_results}
\begin{tabular}{cccc}     
  \hline                   
& PDFI  & raw DAVE4VM & ind. DAVE4VM  \\
& $\bar{\sigma}_{dE_m/dt},\bar{\sigma}_{dH_R/dt}$ & $\bar{\sigma}_{dE_m/dt},\bar{\sigma}_{dH_R/dt}$ & $\bar{\sigma}_{dE_m/dt},\bar{\sigma}_{dH_R/dt}$ \\
Cadence  & [\%] & [\%] & [\%]\\
  \hline
2.25 min & 33, 58 & 30, 20 & 33, 170 \\
11.25 min & 17, 32 & 12, 6.2 & 18, 66 \\
12 min \tabnote{Different data source, and thus different magnetogram noise levels. See Sections 2.1, 2.2 and 2.6 for details.} & 11, 14 & 6.7, 2.8 & 10, 27 \\
2.025 h & 2.4, 5.9 & 9.7, 14 & 3.1, 88 \\
- - (rebin 15x) & 2.5, 11 & 4.4, 7.8 & 2.8, 17 \\
6 h & 1.4, 3.0 & 19, 6.3 & 1.4, 56 \\
- - (rebin 15x) & 0.98, 4.8 & 4.7, 12 & 1.2, 18 \\
12 h & 1.6, 4.4 & 9.4, 12 & 1.6, 81 \\
- - (rebin 15x) & 0.7, 2.7 & 1.4, 1100 \tabnote{This value is spurious. See Section 4.1 for details.} \ & 0.78, 3.2 \\
24 h & 6.5, 5.0 & 17, 23 & 5.1, 10 \\
- - (rebin 15x) & 7.0, 8.6 & 2.2, 5.1 & 7.9, 0.17 \\
  \hline
\end{tabular}
\end{table} 

\section{Results}
	\label{S-E_and_H_inj_results}
	
We employ Eqs. \ref{Eq-E_injection} and \ref{Eq-H_injection} to estimate the total photospheric energy and helicity injection for each of our electric field and magnetogram time series of various cadence and spatial resolutions. Furthermore, as described in Section \ref{S-error_analysis}, we estimate error bars for $E_m(t)$ and $H_R(t)$ at each time $t$. The sections below illustrate the findings for each of our three electric field inversion methods: PDFI as well as the raw and inductive DAVE4VM methods. 

\subsection{PDFI estimates}
	\label{S-PDFI_E_and_H_inj}
	
\begin{figure}[htb]  
    \centerline{\includegraphics[width=0.9\textwidth, trim = 0.0cm 0.75cm 0cm 0.5cm]{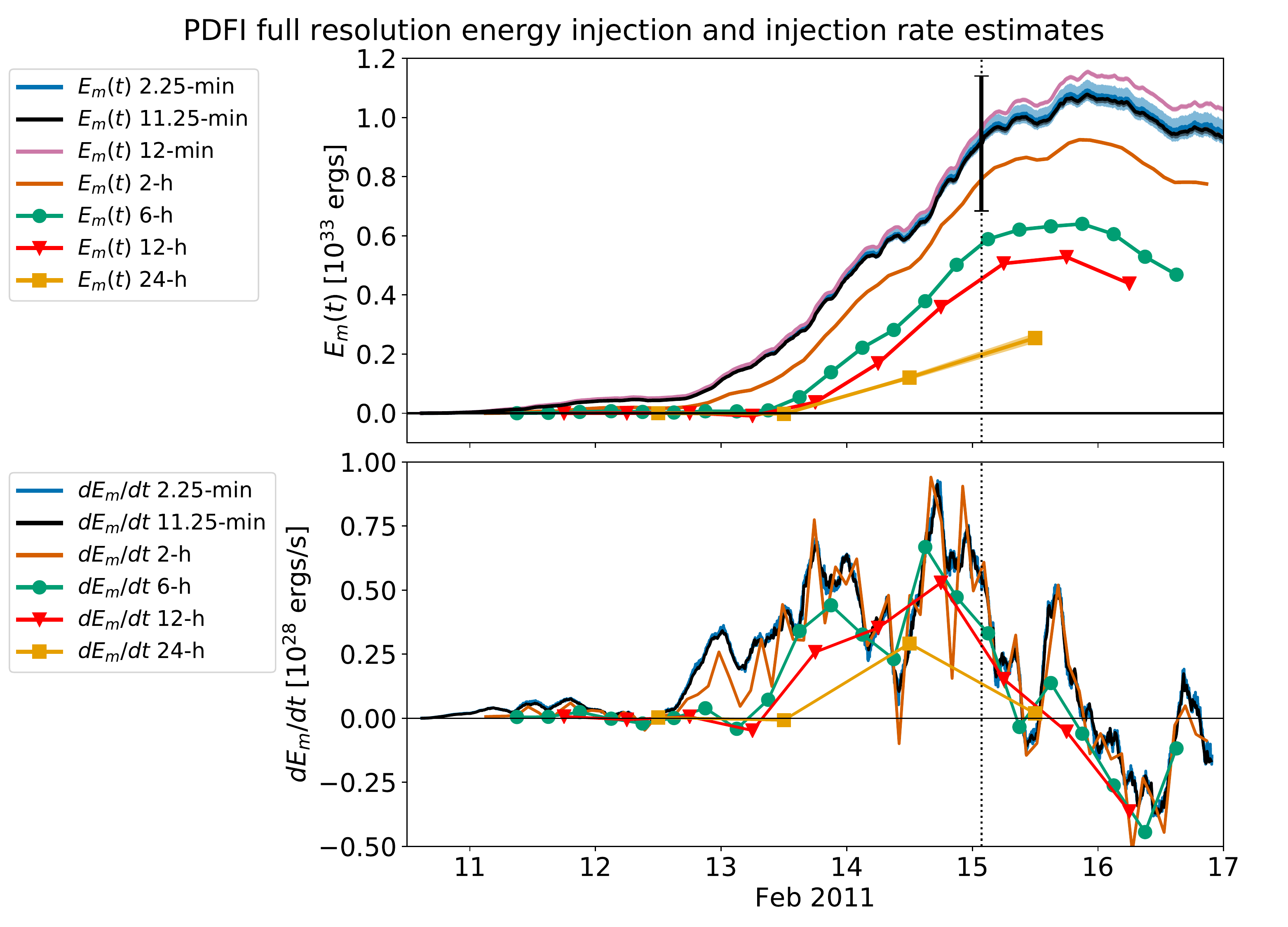}}  
   \caption{Energy $E_m(t)$ injections (upper panel) and energy injection rates $dE_m/dt$ (lower panel) for NOAA AR 11158 derived from the PDFI electric field estimates with variable cadence of the input data. Noise-related error bars of the injection estimates are shown by the shaded regions surrounding the curves in the upper panel. Note that the injection rate curves with cadence $\leq$12 minutes have been smoothed in time using a boxcar of 4 hours to better bring out the temporal trends. Vertical dotted line indicates the strongest X2.2 class flare in AR 11158, and the horizontal solid lines indicate the zeros of the y axes. The black error bar at the time of the flare in the upper panel illustrates the combined method- and noise-related errors of the PDFI energy injection estimate for the 11.25-minute case (see text for details).} 
   \label{F-Em_fluxes_PDFI_full_res}
\end{figure}

\begin{figure}[htb]  
    \centerline{\includegraphics[width=0.9\textwidth, trim = 0.0cm 0.75cm 0cm 0.5cm]{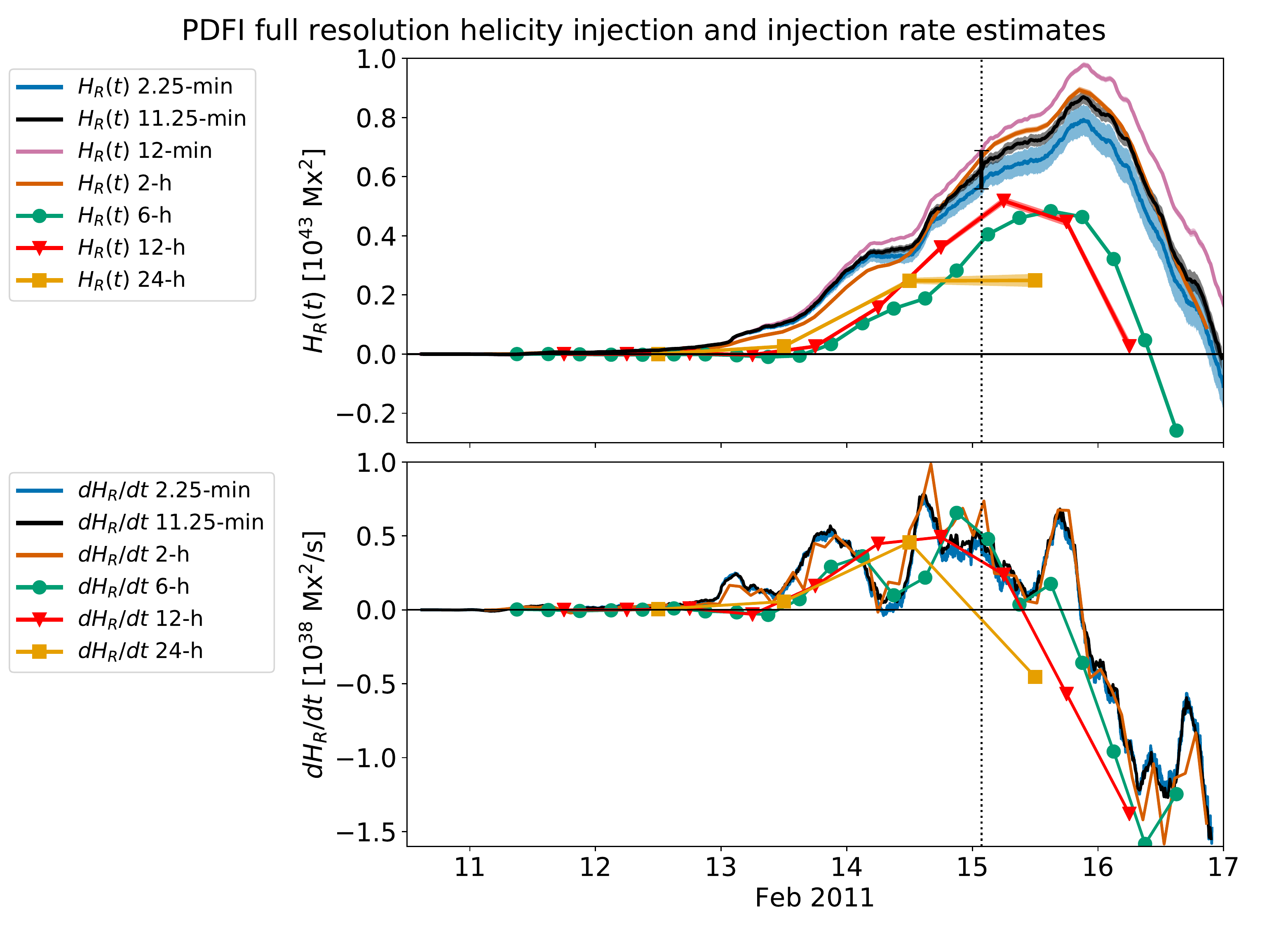}}  
   \caption{Same as Figure \ref{F-Em_fluxes_PDFI_full_res}, but now for the helicity injection $H_R(t)$ and the helicity injection rate $dH_R/dt$.}
   \label{F-HR_fluxes_PDFI_full_res}
\end{figure}

Figures \ref{F-Em_fluxes_PDFI_full_res} and \ref{F-HR_fluxes_PDFI_full_res} illustrate the energy and helicity injections (upper panels) and the injection rates (lower panels) derived from the PDFI electric fields (Section \ref{S_einv_PDFI}) for all cadences with full spatial resolution over the interval February 10, 14:48 -- 17, 00:00 UT, 2011. The time evolution of the energy and helicity injections in NOAA AR 11158 is discussed in detail e.g. by \citet{Liu2012}, \citet{Kazachenko2015} and \citet{Lumme2017}, so we omit most of the discussion in this work. However, we briefly classify the basic phases of the evolution within our analysis interval: the emergence of the active region on Feb 10 22:00 UT \citep[as defined in the HMI SHARP data product,][]{Bobra2014}, after which a time interval of slow magnetic flux emergence and energy and helicity injection continues until Feb 12 $\sim$18:00 UT. After this time a period of strong flux emergence begins \citep[see Figure 4 in][]{Lumme2017}, accompanied with enhanced energy and helicity injections that continue until an X2.2 flare occurs on Feb 15 01:44 UT (dotted vertical line in the plots). After the flare the PDFI energy injection saturates, with the 11.25-minute estimate reaching $E_m(t) \sim 1 \times 10^{33}$ ergs, whereas the helicity injection continues to increase until Feb 16 $\sim$00:00 UT, with the 11.25-minute estimate reaching $H_R(t) \sim 8.7 \times 10^{42}$ Mx$^2$ after which it begins to steeply decrease. 

The errors arising from the magnetogram noise, computed using the Monte Carlo approach, are illustrated by the shaded regions around the injection curves in Figures \ref{F-Em_fluxes_PDFI_full_res} and \ref{F-HR_fluxes_PDFI_full_res}, upper panels. In many cases the error bars are vanishingly small ($\lesssim$1\%, see Section \ref{S-error_bars_disc} for further discussion) and thus invisible in the figure. The error bars are visible for the highest cadence, 2.25 minutes (blue curve), and the mock nominal HMI cadence of 11.25 minutes (black curve), which have the largest errors also in the injection rates for the PDFI estimates (Table \ref{T-error_results}, second column). Consequently, the energy estimates for these cadences are consistent within error bars ($\pm$3\%) at the time of the X-class flare. The difference in the 2.25- and 11.25-minute helicity injection estimates is slightly larger, 8\%, which exceeds the combined error bars by 20\% (in the combined error bars the errors of both estimates are added up in quadrature). The nominal 12-minute estimate (purple curve) has a different data source (Sections \ref{S-proc_magnetograms} and \ref{S-sampling_data}), different magnetogram noise characteristics (Section \ref{S-error_analysis}), as well as different $\sigma_{FLCT}$ parameter (Table \ref{T-optimization_results}), and thus results in different injections and error bars, when compared to the mock nominal cadence of 11.25 minutes (6\% and 10\% larger in $E_m$ and $H_R$ at the time of the X-class flare, both above noise-related error bars). Our 12-minute PDFI estimates differ slightly from the results of \citet{Kazachenko2015}, who used similar input data for this active region: our energy and helicity injection estimates are smaller than theirs by 9\% and 20\% at the time of the X-class flare, respectively. The difference arises both from the updates in the electric field inversion code (Section \ref{S_einv_PDFI} and Appendix \ref{S-app_mods_to_PDFI}), differences in the FLCT optical flow inversion (Section \ref{S-vel_inversion} and Appendices \ref{S-time_dscr_v_and_einv} and \ref{S-mask_in_v_and_einv}) as well as from the differences in data processing and Dopplergram data source (Section \ref{S-proc_magnetograms}).

Although some of the injection estimates discussed above have differences larger than the noise-related error bars, all 2.25-min, 11.25-min and 12-min cases are within the method-related errors of the PDFI method estimated from the ANMHD tests (see Introduction, Section \ref{S-Introduction}), which are: 25\% for the energy injection and 10\% for the helicity injection \citep{Kazachenko2014,Kazachenko2015}. The combined method- and noise-related errors for the 11.25-minute estimates are illustrated by the black error bars in Figures \ref{F-Em_fluxes_PDFI_full_res} and \ref{F-HR_fluxes_PDFI_full_res}, upper panels. However, one should note that we follow here the approach of \citet{Kazachenko2015} and employ the maximal errors from \citet{Kazachenko2014}. Since the errors depend on the viewing angle (i.e. the angle between the $z$ direction and the average LOS direction over the active region patch), at smaller viewing angles the errors are smaller: e.g. for angles $<$20$^{\circ}$ the error in the energy flux drops to 5\%, while the error in helicity flux remains at 10\%. Moreover, even though \citet{Kazachenko2014} define these errors for the injection rates ($dE_m/dt$, $dH_R/dt$) they are not added up in quadrature in the time integration (as in Eq. \ref{Eq-error_prop_in_tint}) since we consider these errors to be systematic in nature, and thus use them directly also in $E_m(t)$ and $H_R(t)$. In other words, the method-related error may result in a systematic under/overestimation of injection rates, which would be then directly visible as an equal relative under/overestimation in the total injections.

When the cadence is lowered to $\geq$2 hours (orange, green, red and yellow curves) we see a gradual decrease in the total energy injection as a function of $\Delta t$ with accumulated energy at the time of the X-class flare dropping monotonically from 86\% to 22\% of the 11.25-minute mock nominal case over the cadences of 2 to 24 hours. For helicity we see similar but less monotonic decrease with accumulated helicities ranging from being 7\% larger (2-hour cadence) to 60\% smaller (24-hour cadence) when compared to the 11.25-minute mock nominal case. Interestingly, for both energy and helicity injection, the 2-hour injection estimate is very close to the high-cadence estimates and well consistent with the 11.25-minute estimate within the combined noise- and method-related error bars. However, the datasets with cadences $\geq$6 hours underestimate the energy and helicity injections beyond all error bars.

When we look at the rate of changes $dE_m/dt$ and $dH_R/dt$ (Figures \ref{F-Em_fluxes_PDFI_full_res} and \ref{F-HR_fluxes_PDFI_full_res}, lower panels), we see reasonably consistent trends over all cadences. For example, the 2.25-minute to 2-hour cases are very similar, and even the 12-hour case approximates well the trends of the higher-cadence curves (where the highest $\leq $12-minute cadences are smoothed using a 4-hour boxcar window to better discern the trends). It has been reported that the highest 2.25-minute cadence allows solar p-mode (5-minute) oscillations to pass over to the $dE_m/dt$ and $dH_R/dt$ injection rates (X. Sun, private communication). We also confirm spikes at $\sim$5 minutes in the Fourier power spectra for both. However, this signal does not propagate notably to the time-integrated quantities $E_m(t)$ and $H_R(t)$, so we will neglect it henceforth.

As explained in Section \ref{S-sampling_data} due to the significant undersampling at the lowest cadences ($\geq$2 hours) we recomputed the energy and helicity injections also from the 15-times rebinned magnetogram series. When it comes to the PDFI results this rebinned version brought very little new features to the energy injection curves of Figure \ref{F-Em_fluxes_PDFI_full_res}, and produced significant loss of helicity injection signal, most dramatically for 2-hour data, which dropped by $\sim$90\% (see Supplementary Figure \ref{F-Em_HR_fluxes_PDFI_rebin}).

\subsection{Raw DAVE4VM electric field estimates}
	\label{S-raw_DAVE4VM_E_and_H_inj}

\begin{figure}[htb]  
    \centerline{\includegraphics[width=0.9\textwidth, trim = 0.0cm 0.75cm 0cm 0.5cm]{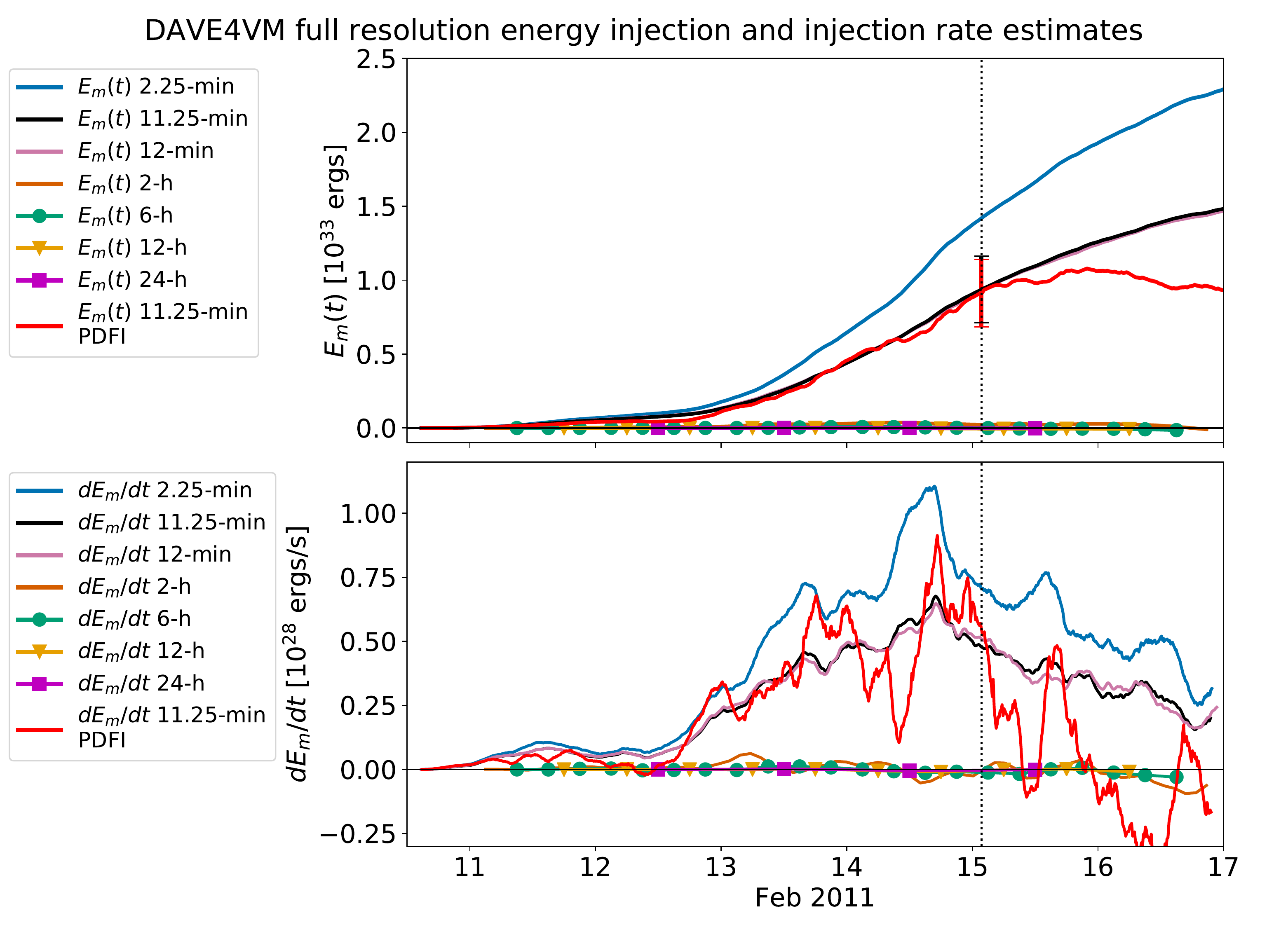}}  
   \caption{Same as Figure \ref{F-Em_fluxes_PDFI_full_res} but now using the raw DAVE4VM electric field for deriving the energy injection estimates. The 11.25-minute PDFI estimate from Figure \ref{F-Em_fluxes_PDFI_full_res} is also plotted for reference (red curve), and the combined method and noise-related error bars are plotted for both 11.25-minute raw DAVE4VM and PDFI estimates (black and red bars) at the time of the X-class flare (vertical black dotted line).} 
   \label{F-Em_fluxes_DAVE4VM_full_res}
\end{figure}

\begin{figure}[htb]  
    \centerline{\includegraphics[width=0.9\textwidth, trim = 0.0cm 0.75cm 0cm 0.5cm]{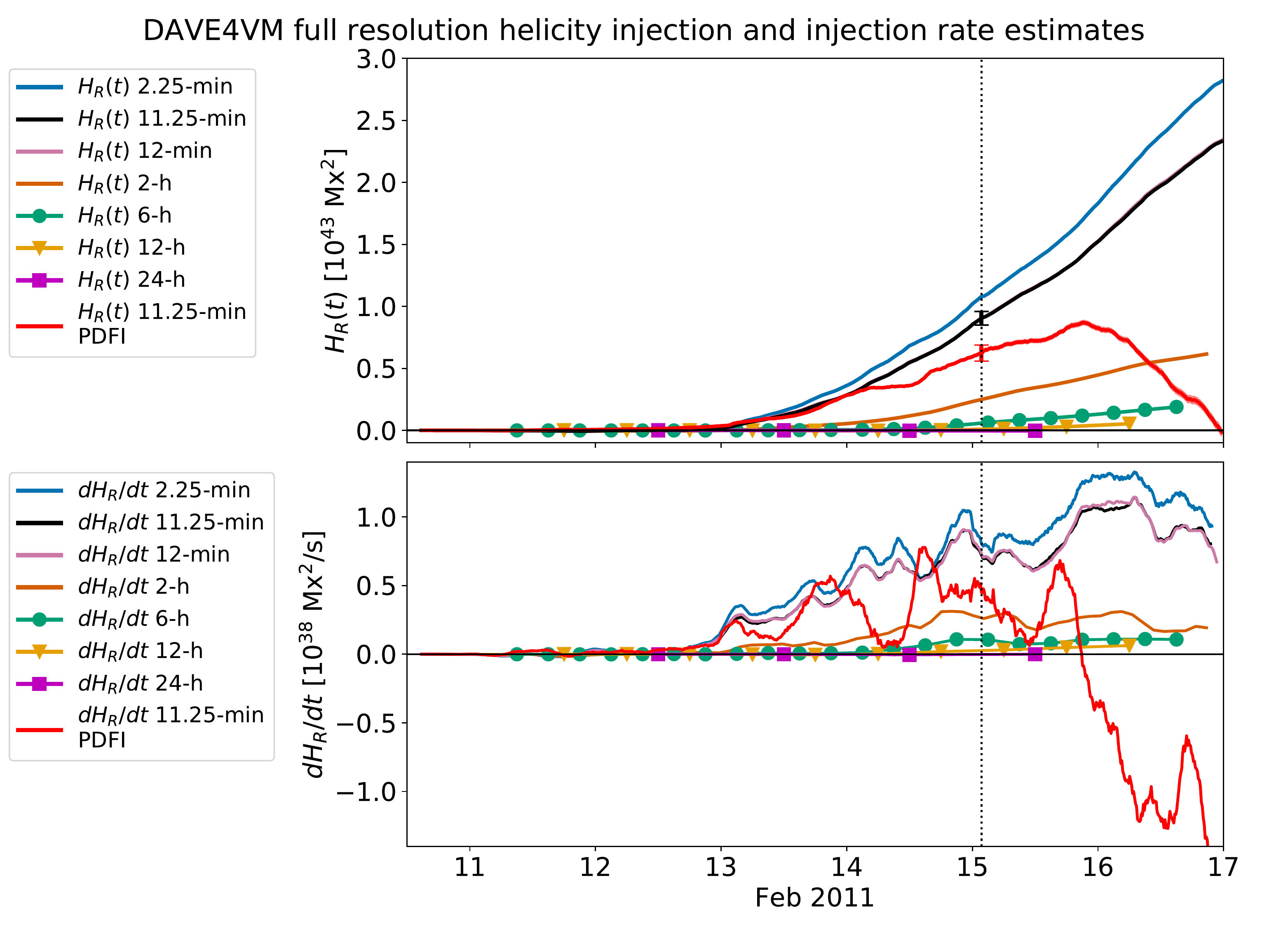}}  
   \caption{Same as Figure \ref{F-Em_fluxes_DAVE4VM_full_res} but now the raw DAVE4VM helicity injections $H_R(t)$ and helicity injection rates $dH_R/dt$ are plotted.} 
   \label{F-HR_fluxes_DAVE4VM_full_res}
\end{figure}

Figures \ref{F-Em_fluxes_DAVE4VM_full_res} and \ref{F-HR_fluxes_DAVE4VM_full_res}, upper panels, illustrate the energy and helicity injections computed from the raw DAVE4VM electric field estimate (Section \ref{S_einv_DAVE4VM}). First, we notice that the highest 2.25-minute cadence produces a clearly higher energy injection (blue curve) than the mock nominal 11.25-minute (black curve) and the nominal 12-minute (purple curve, mostly hidden behind the black curve) cases, being $\sim$50\% larger at the time of the X-class flare, whereas the latter two differ $<$1\% from each other. When we look at the energy injection rate $dE_m/dt$ (Figure \ref{F-Em_fluxes_DAVE4VM_full_res}, lower panel), we find that the discrepancy between the 2.25-minute and 11.25-minute cases in the energy injection is visible as systematically larger $dE_m/dt$ values in the 2.25-minute case, whereas the temporal trends are similar. When it comes to the lowest cadences ($\geq$2 hours), the raw DAVE4VM estimate loses practically all of the energy injection signal producing $\lesssim$2\% of the high-cadence energy injection estimates at the time of the flare. 

Both the helicity injection and injection rate curves (Figure \ref{F-HR_fluxes_DAVE4VM_full_res}, upper and lower panel, respectively) of the three highest cadences (2.25, 11.25 and 12 minutes) are more consistent with each other than in the energy injection case, the 2.25-minute total injection being ``only'' $\sim$20\% larger than the 11.25-minute case at the time of the X-class flare. Similarly to the energy injection, the lowest cadences produce very small helicity injection, the 2-hour case (orange curve) yields 28\% and $\geq$6 h cases (green, yellow, and magenta curves) only $\leq$6\% of the 11.25-min injection at the time of the flare.

Due to very similar data processing and inversion scheme our 12-minute raw DAVE4VM energy and helicity injection estimates at the time of the X-class flare differ $\lesssim$2\% from the results of \citet{Lumme2017} for AR 11158. \citet{Lumme2017} provide further comparison to other raw 12-minute DAVE4VM estimates for AR 11158 by \citet{Liu2012} and \citet{Tziotziou2013}.

The noise-related error estimates are vanishingly small ($<$1\%) for the raw DAVE4VM estimates (shaded regions around each curve are practically invisible in Figures \ref{F-Em_fluxes_DAVE4VM_full_res} and \ref{F-HR_fluxes_DAVE4VM_full_res}, upper panels), but the method-related error bars from the ANMHD-tests are again significantly larger: 24\% and 6\% for the energy and helicity injections, respectively \citep{Schuck2008,Kazachenko2015}. They are plotted for the 11.25-minute estimates in Figures \ref{F-Em_fluxes_DAVE4VM_full_res} and \ref{F-HR_fluxes_DAVE4VM_full_res} -- combined with the noise-related error bars --  as black error bars at the time of the X-class flare.

When comparing the 11.25-minute PDFI reference curve (red curve) and the raw DAVE4VM result (black curve) for energy injection in Figure \ref{F-Em_fluxes_DAVE4VM_full_res}, we find similar evolution between the 11.25- and 12-minute DAVE4VM energy injection curves and the PDFI case until the time of the X-class flare (where the difference is 3\%). After the flare the raw DAVE4VM energy injection continues to increase, whereas the rate of change in the PDFI case becomes occasionally negative and the total injection saturates. Consequently, the DAVE4VM estimate is 57\% larger at the end of Feb 16. When it comes to the helicity injection (Figure \ref{F-HR_fluxes_DAVE4VM_full_res}, upper panel), the PDFI and DAVE4VM results diverge already in early Feb 14, the 11.25-min DAVE4VM result being 45\% larger than the PDFI estimate at the time of the flare, which is beyond the combined noise- and method-related error bars. Due to the strong negative $dH_R/dt$ of the PDFI estimate after Feb 16 00:00 UT (Figure \ref{F-HR_fluxes_DAVE4VM_full_res}, lower panel) the PDFI helicity injection is approximately zero at the end of our time series, whereas the raw DAVE4VM estimate has reached a value of 2.6$H_R(t_{flare})$. This inconsistency between the trends is discussed further in Section \ref{S-role_of_nind_disc}.

\begin{figure}[htb]  
    \centerline{\includegraphics[width=0.9\textwidth, trim = 0.0cm 0.75cm 0cm 0.5cm]{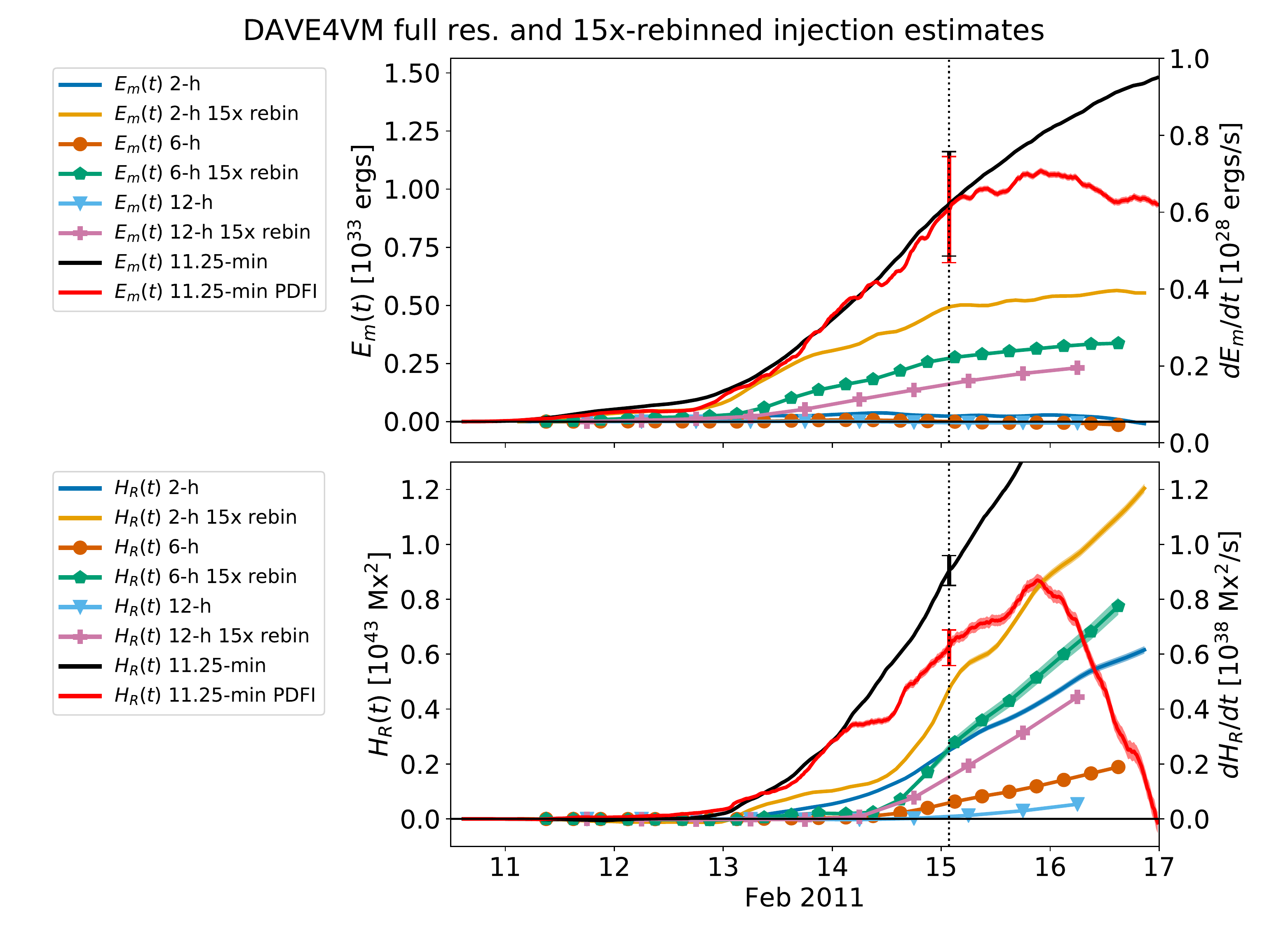}}  
   \caption{Raw DAVE4VM energy $E_m(t)$ and helicity $H_R(t)$ injections derived from the full-resolution and 15-times rebinned estimates for the lowest cadences of 2, 6 and 12 hours. The 11.25-minute full resolution PDFI and raw DAVE4VM estimates are also plotted for reference (red and black curves) with their combined method- and noise-related error bars at the time of the X-class flare (vertical black dotted line).} 
   \label{F-Em_and_HR_fluxes_DAVE4VM_rebin}
\end{figure}

Figure \ref{F-Em_and_HR_fluxes_DAVE4VM_rebin} illustrates the raw DAVE4VM energy and helicity injection estimates derived from both the full resolution and the 15-times rebinned input data. We find a significant increase in most of the estimates after rebinning. More specifically, the 2-, 6- and 12-hour energy injections (blue/yellow, orange/green and light blue/purple curves in Figure \ref{F-Em_and_HR_fluxes_DAVE4VM_rebin}) are increased from approximately zero to 53\%, 29\% and 17\% of the 11.25-minute full resolution reference injection (black curve) at the time of the X-class flare. Similarly, the corresponding helicity injections are increased: 2-hour estimate by a factor of 1.9 reaching 52\% of the full resolution 11.25-minute reference, 6-hour estimate by a factor of 4.4 reaching 28\% of the reference, and 12-hour estimate increasing from approximately zero to 17\% of the reference. Despite the significantly better recovery of the injection signals in the rebinned case, none of the rebinned estimates are consistent with the full resolution high-cadence reference estimates within the combined method- and noise-related error bars (black error bars in Figure \ref{F-Em_and_HR_fluxes_DAVE4VM_rebin}) at the time of the X-class flare. 

\subsection{Inductive DAVE4VM electric field estimates}
	\label{S-ind_DAVE4VM_E_and_H_inj}
\begin{figure}[htb]  
    \centerline{\includegraphics[width=0.9\textwidth, trim = 0.0cm 0.75cm 0cm 0.5cm]{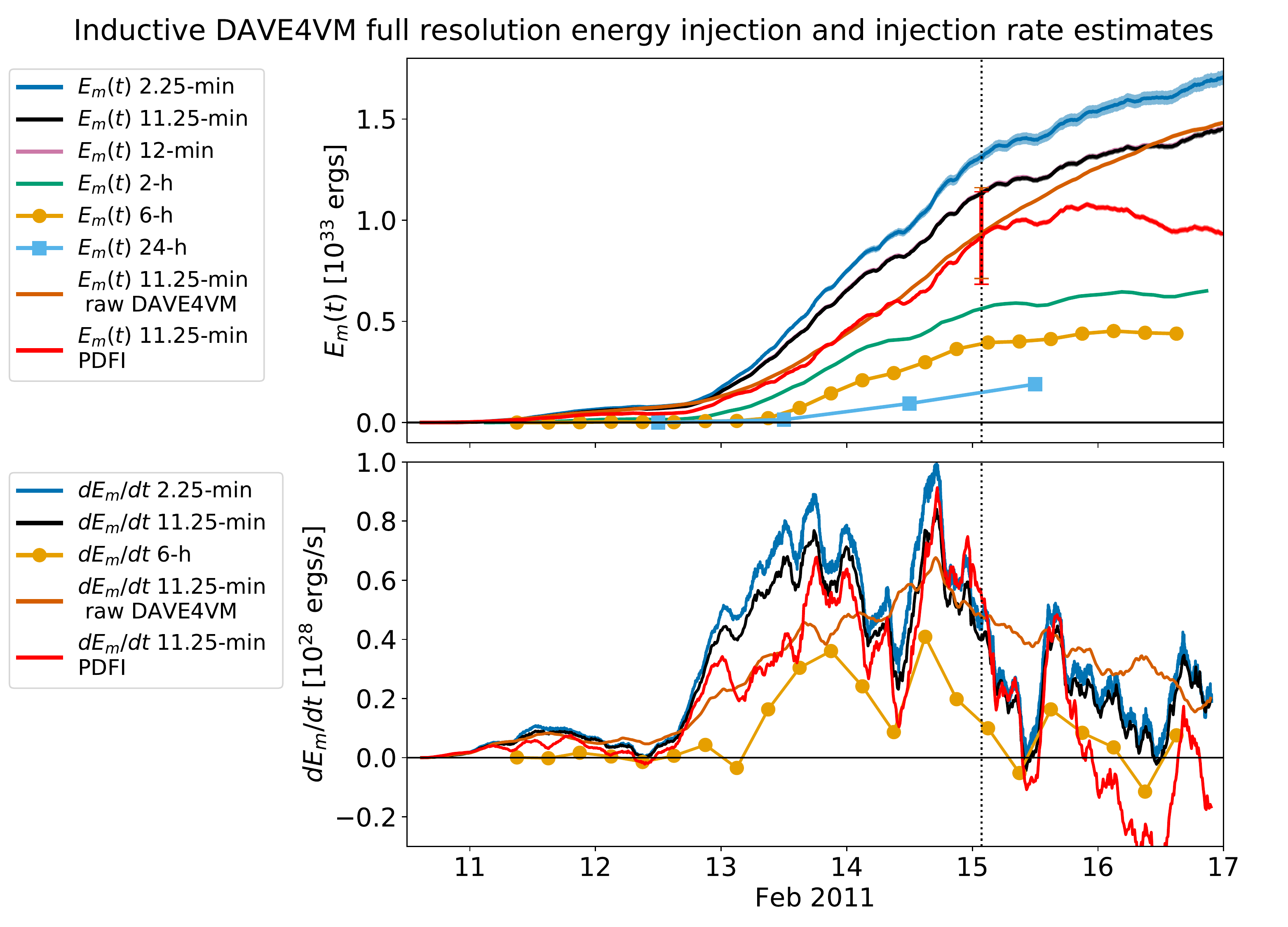}}  
   \caption{Same as Figure \ref{F-Em_fluxes_PDFI_full_res} but now using the inductive DAVE4VM electric field for computing the energy injections. The 11.25-minute PDFI and raw DAVE4VM estimates (red and orange curves) are also plotted for reference with their combined method- and noise-related error bars at the time of the X-class flare (dotted vertical line).} 
   \label{F-Em_fluxes_div_E_DAVE4VM_full_res}
\end{figure}

\begin{figure}[htb]  
    \centerline{\includegraphics[width=0.9\textwidth, trim = 0.0cm 0.75cm 0cm 0.5cm]{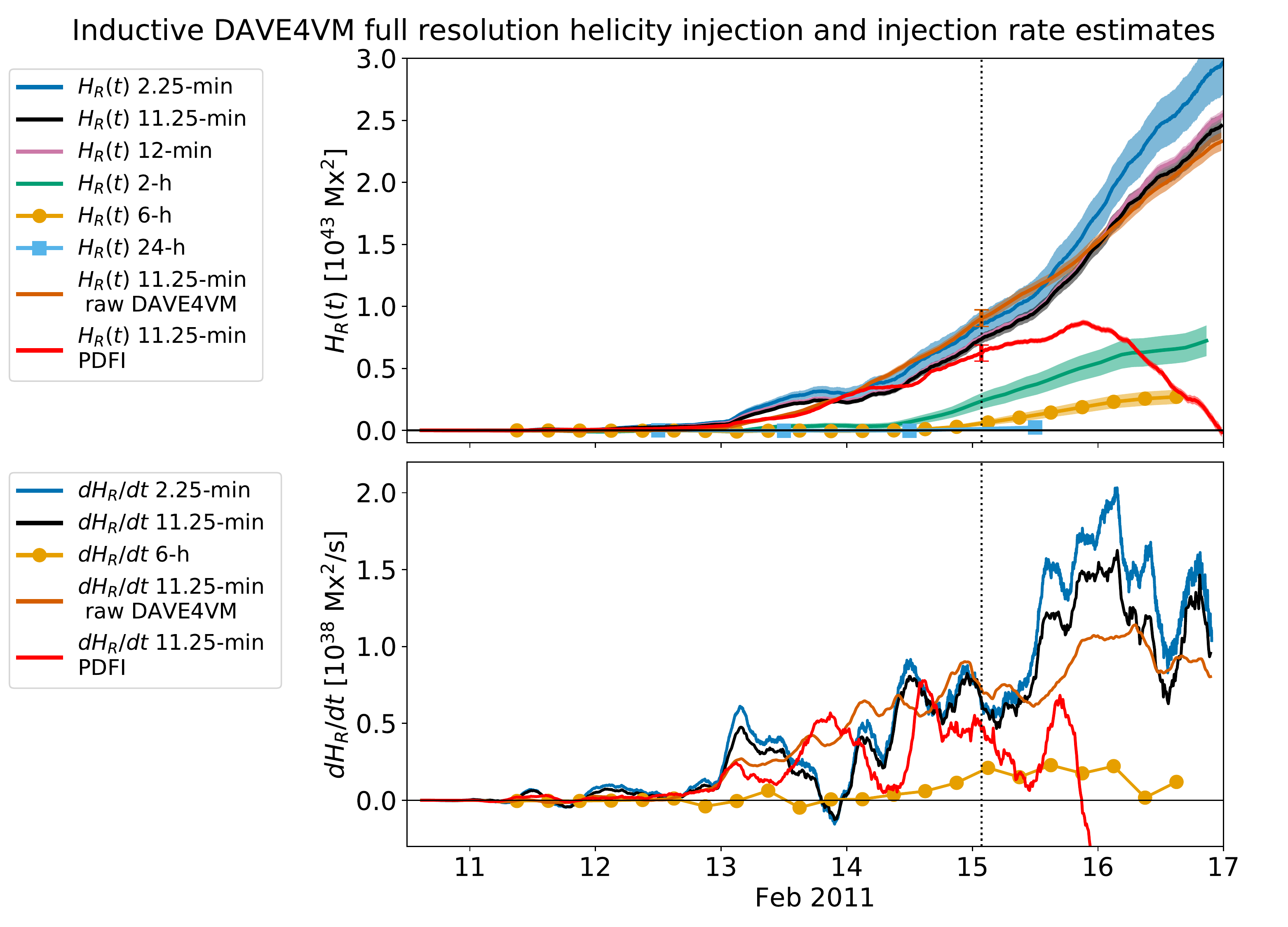}}  
   \caption{Same as Figure \ref{F-Em_fluxes_div_E_DAVE4VM_full_res} but now the inductive DAVE4VM helicity injections $H_R(t)$ and helicity injection rates $dH_R/dt$ estimates are plotted.} 
   \label{F-HR_fluxes_div_E_DAVE4VM_full_res}
\end{figure}

Figures \ref{F-Em_fluxes_div_E_DAVE4VM_full_res} and \ref{F-HR_fluxes_div_E_DAVE4VM_full_res}, upper panels, illustrate the energy and helicity injections computed from the inductive DAVE4VM estimates (Section \ref{S_einv_DAVE4VM}). Similarly to the raw DAVE4VM estimate the 2.25-minute estimate (blue curves) overestimates the energy injection of the 11.25- and 12-minute cases (black and purple curves, which are again mutually consistent, the latter curve being mostly hidden behind the former), but now clearly less, by only 16\% (instead of $\sim$50\% as in the raw case) at the time of the X-class flare. The temporal trends of both $E_m(t)$ and $dE_m/dt$ are very similar over all three cadences. Unlike for the raw DAVE4VM estimate now also the lowest cadences ($\geq$2 hours; green, orange and light blue curves) produce noticeable energy injections reaching $\lesssim$50\% of the 11.25-minute estimate at the time of the X-class flare. However, this arises mostly from the inductive component of the electric field, the DAVE4VM-based non-inductive contribution being vanishingly small.

The 11.25-minute inductive DAVE4VM energy injection estimate (black curve) overestimates the raw DAVE4VM estimate (orange curve) in Figure \ref{F-Em_fluxes_div_E_DAVE4VM_full_res}, upper panel. However, the differences between the inductive and raw estimates as well as the PDFI estimate are within the combined noise- and method-related error bars of the raw DAVE4VM and the PDFI estimates (red/orange bars) at the time of the X-class flare. When it comes to the helicity injections in Figure \ref{F-HR_fluxes_div_E_DAVE4VM_full_res}, upper panel, $H_R(t)$ curves have very similar temporal trends between the raw and inductive DAVE4VM estimates, however, with noticeable systematic differences. The raw estimate is larger than the inductive one by $\lesssim$27\% for cadences $\leq$6 hours at the time of the X-class flare, which are above the method-related error bars of the raw estimate. The inductive DAVE4VM helicity estimate is 18\% larger than the PDFI case, beyond all error bars. 

Unlike in the raw DAVE4VM energy and helicity estimates, the noise-related error bars (indicated by the shaded regions around the curves in Figures \ref{F-Em_fluxes_div_E_DAVE4VM_full_res} and \ref{F-HR_fluxes_div_E_DAVE4VM_full_res}, upper panels) are not completely insignificant for the inductive DAVE4VM estimates, particularly in the case of helicity injection. For example, the relative error in the 2.25-minute energy estimate is 2\%, and the relative errors in the 2.25-minute and 2-hour helicity estimates are 13\% and 27\%, respectively.  

Finally, we tested the effect of rebinning the data for the lowest cadences ($\geq$2 hours) also with the inductive DAVE4VM estimate. In energy injection curves (see Supplementary Figure \ref{F-Em_HR_div_E_DAVE4VM_rebin}, upper panel) we find consistent increases $\sim$20--70\% over all cadences, while the temporal trends remained the same. The effect of the rebinning on the helicity injection is, however, clearly more dramatic. As illustrated by Supplementary figure \ref{F-Em_HR_div_E_DAVE4VM_rebin}, lower panel, the 2-, 6- and 12-hour estimates change their sign in the rebinning yielding negative total injections at the time of the X-class flare. This result is problematic and likely spurious and will be discussed more in detail in Section \ref{S-role_of_nind_disc}.

\section{Discussion}
	\label{S-Discussion}
	
In this section we discuss the findings made in Section \ref{S-E_and_H_inj_results}, their reliability and possible issues. We also compare our results to previous works and discuss some general implications of our results.

\subsection{Further testing of the error estimates}
	\label{S-error_bars_disc}
	
One central finding of our study of the energy and helicity injections in Section \ref{S-E_and_H_inj_results} was that the error bars arising from the noise in the magnetogram data were very small, often $<$1\%, which is clearly less than the errors of $\sim$10\% reported e.g. by \citet{Kazachenko2015}. The small magnitude of the errors arises from the error propagation in the time integration, where the errors of individual injection rates ($dE_m/dt$ and $dH_R/dt$) -- which may be substantial (see Table \ref{T-error_results}) -- are summed up in quadrature, thus reducing the error in the total injections ($E_m(t)$ and $H_R(t)$) roughly by factor of $\sqrt{N}$, where $N$ is the number of times $t_i \leq t$ in the time series. 

Our error estimates have, however, significant uncertainties. One particular issue is the use of constant relative error estimates for the injection rates over the entire time interval, derived for a single representative frame near Feb 14 01:48 UT. Although the frame was chosen carefully, so that AR 11158 has its central magnetic structures already emerged at this time while the region is still experiencing strong flux emergence and energy and helicity injection, there is no guarantee that the relative error estimate is accurate when used over the entire analysis interval. Thus, to validate this approach we conducted further Monte Carlo tests. Instead of studying a single frame in the time series, we perturbed the magnetic field data with the same Gaussian noise we used for the representative frame (Section \ref{S-error_analysis}) at all times, and computed the total energy and helicity injections for each Monte Carlo realization. Due to our large number of 33 data series we, however, could not afford to do as many realizations as for the representative frame, and thus we lowered the number of realizations from 199 to 10 for the full resolution cases, and 20 for the rebinned, low cadence cases. This small number of MC realizations makes the statistical uncertainty of these additional error estimates very high, and thus the quantitative values retrieved should be treated only as rough estimates. However, as illustrated below the comparison between this additional MC test and the default error estimates still presented a reasonable test for the default errors and revealed also interesting systematic effects.

\begin{figure}[htb]  
    \centerline{\includegraphics[width=\textwidth, trim = 0.0cm 0.0cm 0cm 0.0cm]{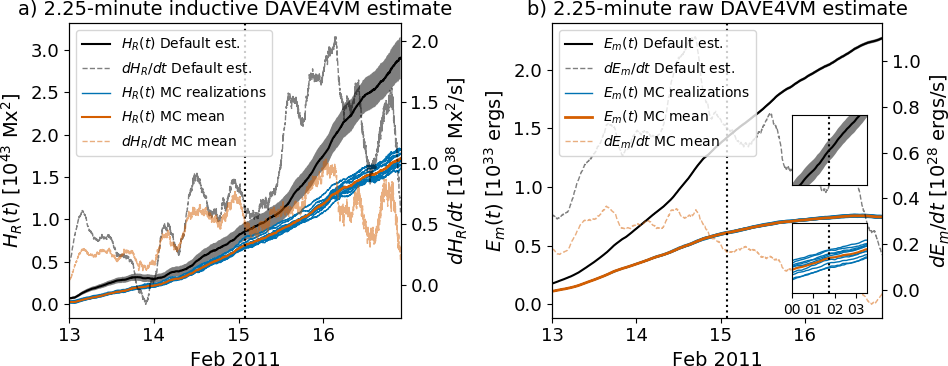}}  
   \caption{Comparison between the unperturbed energy and helicity injection curves (black curves), their default error bars (gray shaded region around the black curves), and the 10 perturbed Monte Carlo realizations (blue curves), and the mean over the Monte Carlo realizations (orange curve). We have plotted also the injection rates for the unperturbed case and the MC mean (black/orange dashed curves). Panel a) shows the results for the 2.25-minute inductive DAVE4VM estimate. Panel b) contains the 2.25-minute raw DAVE4VM energy injection estimates, where the smaller panels, with equal y-axis height (in ergs), zooms into the time of the X-class flare on Feb 15 01:44 UT (vertical black dotted line) to better show the very small $\lesssim$1\% default error bars (upper panel) and the spread of the MC realizations (lower panel).} 
   \label{F-error_bars_MC_comp}
\end{figure}

Figure \ref{F-error_bars_MC_comp} compares the original unperturbed curve and its default error bars (black curves, and shaded grey region around it) to the 10 Gaussian-perturbed Monte Carlo (MC) realizations (blue curves) and their mean (orange curve) for the 2.25-minute inductive DAVE4VM helicity injection estimate (panel a), and for the 2.25-minute raw DAVE4VM energy injection estimate (panel b).
Panel a) illustrates how the relative error estimate for the injection rate gets lowered when the error is propagated from the injection rate to the time-integrated total injection, the largest relative error of 170\% in $dH_R/dt$ (Table \ref{T-error_results}) dropping to 13\% in the total injection $H_R(t_{X\textrm{-}flare})$. We also see that the 10 MC realizations (blue curves) have a very similar spread around the mean of the MC realizations (orange curve). The standard deviation of the MC realizations w.r.t. the mean is 64\% smaller than the standard deviation derived from our default error bars at the time of the X-class flare, corresponding to a relative error of 6\% w.r.t the MC mean. We get similar results for the raw DAVE4VM estimate in panel b), though the error bars are significantly smaller. As the injection curves for the additional MC realizations have no significant outliers, we assume that it is safe to interpret the spread around the MC mean as an approximation of the true error despite the large statistical uncertainty arising from the small number of only 10 MC realizations.

We find similar results as above also for all the other time series in our study: the default error bars and the spread of the MC realizations are mostly very similar, the absolute value of the latter varying between 10\% to 370\% of the former with median of 60\%.\footnote{In the case of the rebinned 12-hour raw DAVE4VM helicity injection estimate our default error estimate failed completely, giving a relative error of 1100\% for the injection rate, which then propagated to the helicity injection giving a relative error $\gtrsim$100\%. This spurious result arises from the very small helicity injection at the representative frame for this particular series, which makes the $\mu$ in Eq. \ref{Eq-rel_sigma_error} small, and thus the $\bar{\sigma}_{dH_R/dt}$ unrealistically large. Due to this issue error bars are omitted from the 12-hour rebinned helicity injection estimate in Figure \ref{F-Em_and_HR_fluxes_DAVE4VM_rebin}, lower panel.}

Despite differences between the MC spread and the default error we still find our default error bars performing reasonably well when their absolute values are compared to the \emph{spread of the MC realizations around their mean}, particularly when considering the inherent uncertainties of the default error bars and the large statistical uncertainty in the spread of the small number of additional MC realizations.

However, we find also significant systematic differences between the MC realizations and the default estimate. This can be seen clearly in both Figures \ref{F-error_bars_MC_comp}a and b; the MC realizations produce consistently lower injections, 57\% smaller for the 2.25-minute raw DAVE4VM energy injection estimate, and 19\% smaller for the 2.25-minute inductive DAVE4VM helicity injection estimate at the time of the X-class flare. The consistency of the underestimation over all MC realizations implies that this is not a spurious result arising from the small number of only 10 realizations. These kind of systematic changes are observed for all of our data series, and they are mostly clearly beyond the noise-related error bars, and in some cases (such as Figure \ref{F-error_bars_MC_comp}) even beyond the method-related error bars of the DAVE4VM and PDFI methods (see Section \ref{S-E_and_H_inj_results}). 

These consistent changes across the datasets show that all of our electric field inversion methods and consequently the energy and helicity injection estimates have systematic responses to the Gaussian perturbations of the input magnetogram data, and that these systematic differences are of a significant nature. In addition to the (unspecified) method-related causes for the observed systematic responses, also the Gaussian noise perturbations that we add to the magnetograms (Section \ref{S-error_analysis}) may introduce biases themselves, mostly because their width is constant in time and space, and chosen to overestimate all temporal variations in the true noise levels in the magnetic field components \citep[see Figure 2 in][]{Kazachenko2015}. Consequently, when perturbing the magnetograms with Gaussian noise, we actually increase the noise levels of the data. After recognizing this, it is not a surprise to find systematic effects, particularly effects where part of the signal seems to be lost as in the Figures \ref{F-error_bars_MC_comp}a and b. Clearly a more comprehensive study on the effect of the noise is required so that more realistic and time-dependent magnitudes for the perturbations are used with larger number of MC realizations than what used here. Until the cause behind these systematic effects is unraveled, care should be taken when deriving energy and helicity injection estimates from input datasets with variable noise levels, as the noise may have significant systematic effects on the result.

Since the systematic effects in the injections only separate the unperturbed cases (black curves in Figure \ref{F-error_bars_MC_comp}) from the 10--20 perturbed MC realizations (blue curves in Figure \ref{F-error_bars_MC_comp}), they do not negate our original finding that the error propagated into the total time-integrated injections is small, which was recovered both for the unperturbed estimate with its default error bar and for the spread of the MC realizations around their mean. The following conclusion that the noise-related variations in the total injections are actually often very small ($<$1\% for the 12-minute estimates, which are most consistent with previous studies) clearly puts a stronger emphasis on other sources of uncertainties when comparing different results. These include the method-related errors within PDFI and DAVE4VM \citep{Kazachenko2014,Schuck2008} already discussed in Section \ref{S-E_and_H_inj_results} as well as the effects of data processing such as Dopplergram calibration \citep[the effect of which was shown to be significant by][]{Kazachenko2014} as well as tracking speed of the active region, noise masking threshold and the azimuth disambiguation in the magnetograms \citep[which all were shown to produce significant effects by][]{Lumme2017}.

\subsection{Inductive and non-inductive contributions to the energy and helicity injections}
	\label{S-role_of_nind_disc}

Recently, there have been active discussions on the importance of the non-inductive electric field, and many studies highlight its crucial role in realistically producing magnetic helicity and energy injections, and the related eruptive activity in the corona  \citep{Cheung2012,Kazachenko2014,Mackay2014,Lumme2017,Pomoell2019}. In this section we quantify the mutual importance of the inductive (constrained by Faraday's law and input magnetograms) and non-inductive contributions (constrained by velocity estimates and Ohm's law) in our injection estimates.

\begin{figure}[htb]  
    \centerline{\includegraphics[width=0.9\textwidth, trim = 0.0cm 0.75cm 0cm 0.75cm]{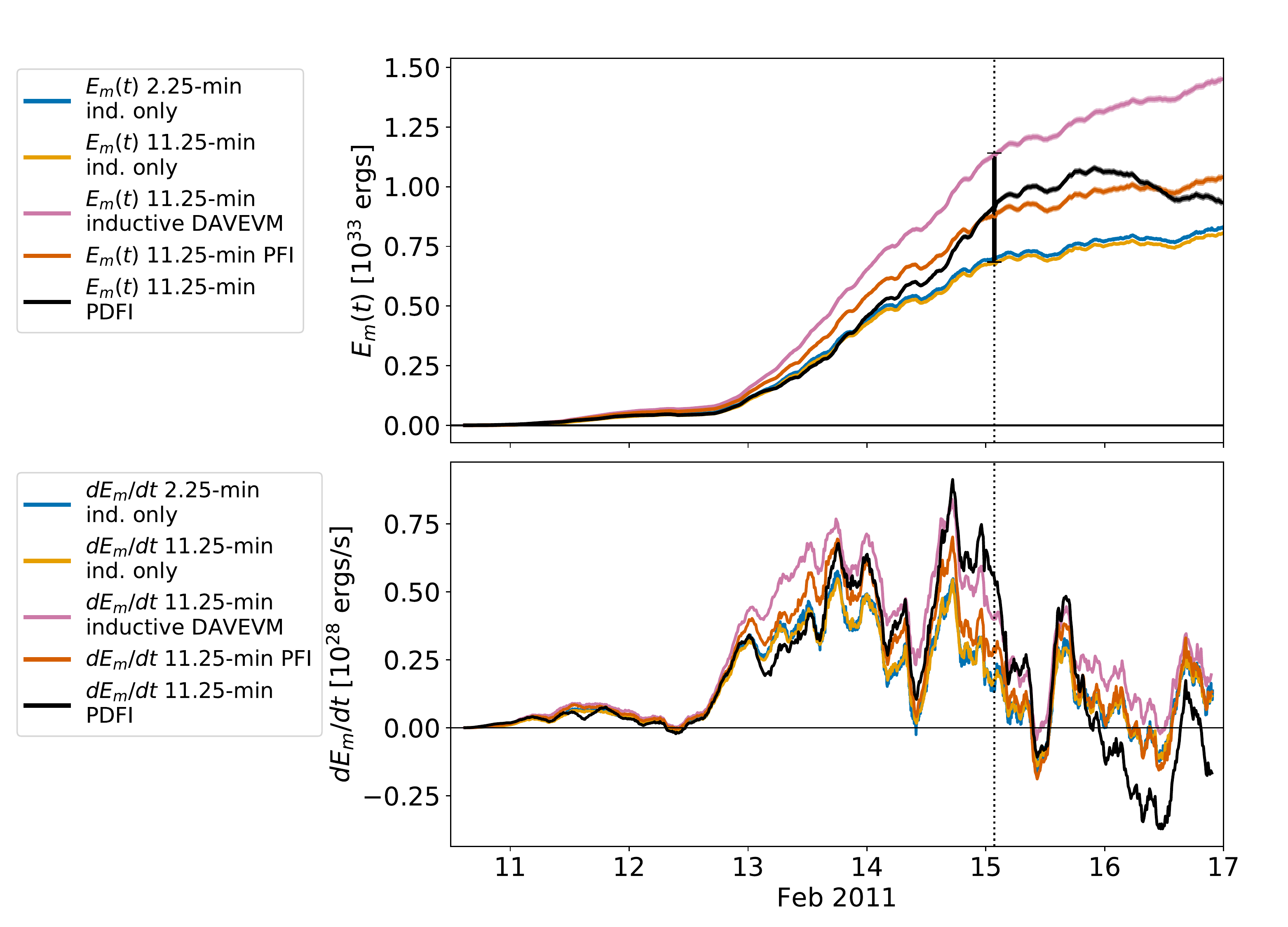}}  
   \caption{Energy injection $E_m(t)$ (upper panel) and injection rate $dE_m/dt$ (lower panel) for the high-cadence 2.25- and 11.25-minute cases derived only from the inductive electric field component, as well as the 11.25-minute estimates derived using the PDFI, inductive DAVE4VM and PFI (PDFI estimate without the Dopplergram contribution) methods. Combined method- and noise-related error estimates are plotted for the 11.25-minute PDFI estimate (black curve) at the time of the X-class flare (vertical black dashed dotted line).}
   \label{F-Em_fluxes_div_ED_PDFI_PFI_and_P}
\end{figure}

\begin{figure}[htb]  
    \centerline{\includegraphics[width=0.9\textwidth, trim = 0.0cm 0.75cm 0cm 0.75cm]{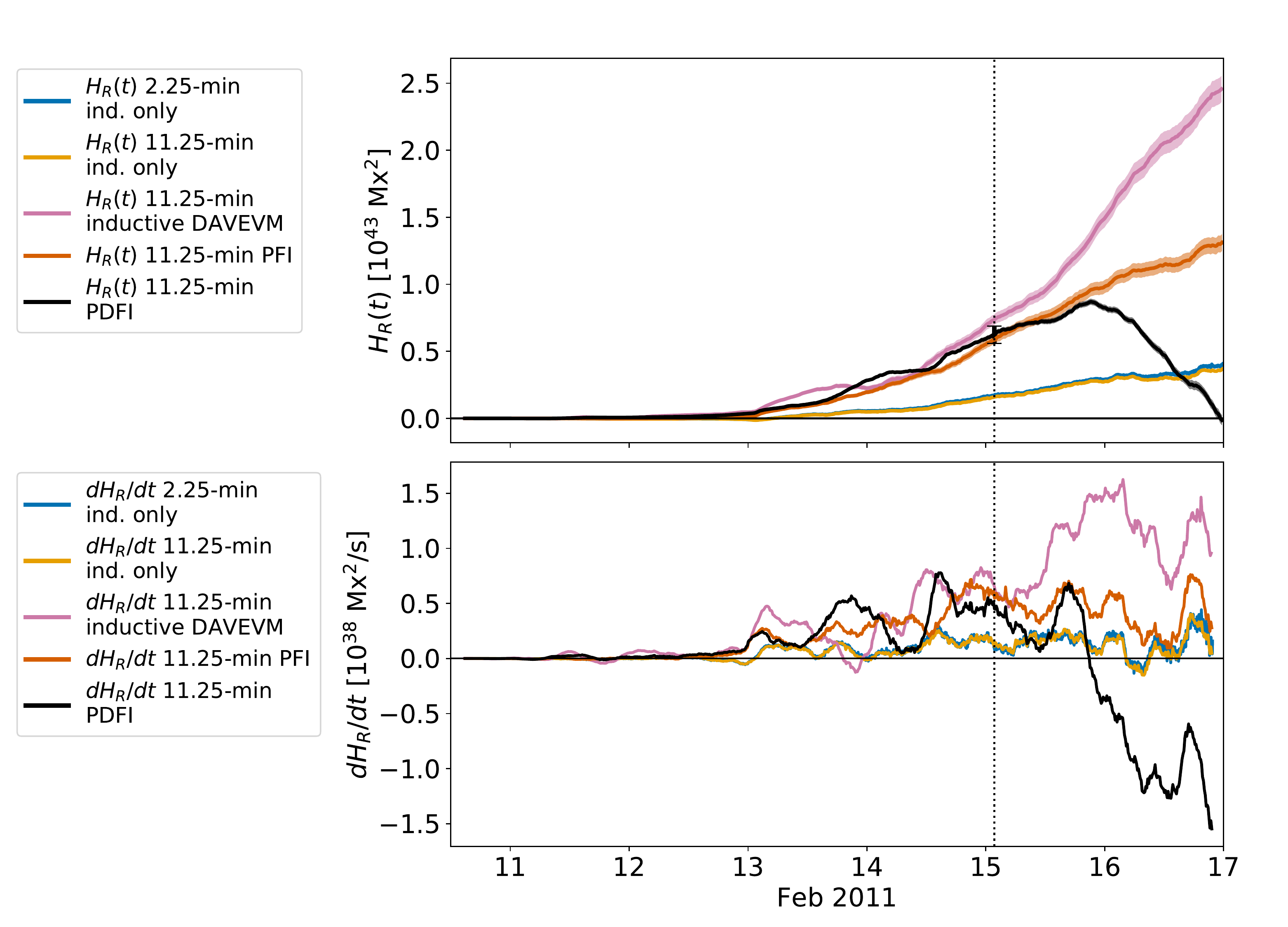}}  
   \caption{Same as Figure \ref{F-Em_fluxes_div_ED_PDFI_PFI_and_P}, but now the role of the inductive component in the helicity injections $H_R(t)$ and injection rates $dH_R/dt$ is illustrated.}
   \label{F-HR_fluxes_div_ED_PDFI_PFI_and_P}
\end{figure}

Figures \ref{F-Em_fluxes_div_ED_PDFI_PFI_and_P} and \ref{F-HR_fluxes_div_ED_PDFI_PFI_and_P}, upper panels, illustrate the significance of the 11.25-minute inductive electric field contribution to the energy and helicity injections (yellow curves) with the full PDFI estimate (black curves) and inductive DAVE4VM estimate (purple curves). First, we notice that the inductive contribution is significant for both inductive DAVE4VM and PDFI energy injection estimates. At the time of the X-class flare the inductive contribution is 60\% of the former and 74\% of the latter. The difference between the inductive contribution and the total PDFI energy injection is almost within the maximal method-related error bars of the method (25\%, see Section \ref{S-PDFI_E_and_H_inj}). The large inductive contribution in the PDFI estimate for NOAA AR 11158 is in contrast with the results of \citet{Kazachenko2014}, who found that the inductive component produced only 40\% of the total PDFI energy injection rate in the ANMHD synthetic data test. This highlights the differences between the properties of the NOAA active region 11158 and the active-region-like system in the ANMHD data, and illustrates the need for testing the electric field inversion methods in other cases beside these two.

When it comes to the helicity injection (Figure \ref{F-HR_fluxes_div_ED_PDFI_PFI_and_P}, upper panel), the inductive component has a significantly smaller contribution compared to the energy injection: at the time of the X-class flare the inductive component constitutes only 22\% and 26\% of the 11.25-minute inductive DAVE4VM and PDFI estimates, respectively. Again our results are in contrast to \citet{Kazachenko2014}, who found the inductive contribution in the helicity injection rate of the ANMHD test to be 55\%, twice larger than in our case. 

Since the inductive contribution constitutes only 20--60\% of the injections for the inductive DAVE4VM method, the non-inductive component extracted from the raw DAVE4VM electric field clearly introduces significant contributions (40--80\%). This result is not directly obvious due to the fact that the method is based on finding a solution that is as inductive as possible.

We plot the inductive contribution for both the 2.25- and 11.25-minute input data in Figures \ref{F-Em_fluxes_div_ED_PDFI_PFI_and_P} and \ref{F-HR_fluxes_div_ED_PDFI_PFI_and_P} (blue and yellow curves) to illustrate that increasing the cadence from 11.25 minutes to the highest 2.25-minute case has a very small effect ($\lesssim$6\%) on both injections. Since increasing the cadence from 11.25 to 2.25 minutes does not cause any significant changes in the inductive energy injection, the large increase in the DAVE4VM-based energy injection estimates between the 11.25- and 2.25-minute cadences (Figures \ref{F-Em_fluxes_DAVE4VM_full_res} and \ref{F-Em_fluxes_div_E_DAVE4VM_full_res}, upper panels) must arise from contributions external to the inductive electric field. These contributions can be divided in two using a 2-dimensional Helmholtz decomposition of the the horizontal electric field \citep[similar to Eq. 7 of][]{Schuck2008} (the vertical, $z$ component of the electric field does not contribute to the injections):
\begin{equation}
\label{Eq-Helmholtz_decomp_2D}
\vec{E}_h = -\nabla \phi \times \hat{\vec{z}} - \nabla_h \psi = -\nabla \times \phi\hat{\vec{z}} - \nabla_h \psi
\end{equation}
We can immediately see that the curl-free part $-\nabla_h \psi$ of this decomposition corresponds to the horizontal components of the non-inductive electric field from Eq. \ref{Eq-ind_nind_decomp}. The divergence-free part $-\nabla \times \phi\hat{\vec{z}}$, on the other hand, defines the inductivity properties of the electric field in the $z$ direction:
\begin{equation}
\label{Eq-div_free_phi_Faraday}
(\nabla \times \vec{E}) \cdot \hat{\vec{z}} = (\nabla \times \vec{E}_h) \cdot \hat{\vec{z}} = -\left[\nabla \times (\nabla \times \phi\hat{\vec{z}})\right] \cdot \hat{\vec{z}} = \nabla_h^2 \phi \ \left(= -\frac{\partial B_z}{\partial t}\right),
\end{equation} 
where the last equality is fulfilled only if the electric field is perfectly inductive. This is the case for the PDFI and inductive DAVE4VM estimates (Eqs. \ref{Eq-ind_nind_decomp} and \ref{Eq-ind_DAVE4VM_e_field}), for which the divergence-free part corresponds to the horizontal inductive electric field $-\nabla \times \phi \hat{\vec{z}} = \mathbf{E}_I^h$. However, since the raw DAVE4VM estimate is poorly inductive (as discussed already in Section \ref{S_einv_DAVE4VM} and as indicated by the metrics in Appendix \ref{S-opt_res_FLCT_and_D4VM}, Table \ref{T-optimization_results}), the divergence-free component of the horizontal raw DAVE4VM electric field $-\nabla \times \phi \hat{\vec{z}}$ is different from the inductive electric field and does not fulfill the normal component of Faraday's law in Eq. \ref{Eq-div_free_phi_Faraday}.

Now we can compare the energy injection estimates derived from the divergence- and curl-free parts of the horizontal DAVE4VM-based electric fields and how they respond to variable cadence. First, the horizontal inductive electric field $\mathbf{E}_I^h$, which is also the divergence-free part of the inductive DAVE4VM electric field in Eq. \ref{Eq-Helmholtz_decomp_2D}, produces only a minimal increase between the 11.25- and 2.25-minute estimates (3\%, $2.2 \times 10^{31}$ ergs, at the time of the X-class flare). Since the 2.25-minute inductive DAVE4VM estimate is 16\% ($1.8 \times 10^{32}$ ergs) larger than the 11.25-minute case, almost all of this (90\%, $1.6 \times 10^{32}$ ergs) must arise from the curl-free, non-inductive part of the inductive DAVE4VM electric field $-\nabla_h \psi$ (see Eq. \ref{Eq-ind_DAVE4VM_e_field}). The horizontal raw DAVE4VM electric field contains the same curl-free part as the inductive DAVE4VM estimate $-\nabla_h \psi$, but the divergence-free part $-\nabla \times \phi \hat{\vec{z}}$ is different and poorly inductive. The 2.25-minute raw DAVE4VM energy injection estimate is 52\% ($4.8 \times 10^{32}$ ergs) larger than the 11.25-minute case. This means that the curl- and divergence-free parts together introduce $4.8 \times 10^{32}$ ergs difference between the 2.25- and 11.25-minute raw DAVE4VM energy injections, whereas the curl-free part alone introduces an increase of roughly $1.6 \times 10^{32}$ ergs. The divergence-free part is thus responsible for most ($(4.8-1.6 = 3.2) \times 10^{32}$ ergs, 66\%) of the difference. The increase from the curl-free, non-inductive component alone ($1.6 \times 10^{32}$ ergs) would already be within the combined method- and noise-related error of the 11.25-minute raw DAVE4VM estimate (which is $2.2 \times 10^{32}$ ergs, $\sim$24\%), making the 2.25- and 11.25-minute estimates consistent within these uncertainties. As indicated by Table \ref{T-optimization_results} in Appendix \ref{S-opt_res_FLCT_and_D4VM} the divergence-free part of the 2.25-minute raw DAVE4VM electric field is clearly less inductive than that of the 11.25-minute case, which means that the divergence-free component of the 2.25-minute raw DAVE4VM electric field is clearly different from that of the 11.25-minute case. Thus, it is not a surprise that these produce different energy injections. 

In conclusion, this analysis shows that the majority of the observed difference between the 2.25- and 11.25-minute raw DAVE4VM energy injection estimates arises from the divergence-free part of the raw DAVE4VM electric field and its variable inductivity properties. The reason why the induction equation/Faraday's law is fulfilled only in least-squares sense in DAVE4VM is that when ensuring perfect inductivity, also the changes related to the noise are reproduced, which is deemed undesirable (\citealp{Schuck2008}, see also Section \ref{S_einv_DAVE4VM}). However, nothing guarantees that the electric field derived from such an approximate reproduction of Faraday's law would evolve the magnetic field in a ``mean field sense'' so that the large scale evolution is reproduced without undesirable noise-related evolution. Instead, the poorly inductive raw DAVE4VM electric field evolves the magnetic field further and further away from the actual observations over time. Thus, we conclude that the changes in the energy injection introduced by variable inductivity between the cadences have little physically justifiable basis and arise only from the specific formulation of the method. Consequently, we recommend ensuring the inductivity of the method not only when using DAVE4VM-based electric fields to drive coronal simulations \citep[as noted already by][]{Schuck2008}, but also when deriving the evolutionary energy and helicity injection estimates from the DAVE4VM electric fields. This has not been taken into account in previous evolutionary estimates derived from DAVE4VM velocities \citep[e.g.][]{Liu2012,Tziotziou2013,Liu2016,Lumme2017,Bi2018}.

Another DAVE4VM-related finding we made in Section \ref{S-E_and_H_inj_results} was that the 15-times rebinned low-cadence inductive DAVE4VM estimates produced negative total helicity injection for NOAA AR 11158 at the time of the X-class flare (Supplementary Figure \ref{F-Em_HR_div_E_DAVE4VM_rebin}, lower panel), which is in contrast with all studies of the active region \citep[see][for review]{Kazachenko2015}. As the helicity injection estimates derived only from the inductive electric field component were consistently positive, this signal must arise from the non-inductive, DAVE4VM-based contribution. As we could not isolate any problematic structures in the electric field or helicity flux density maps (see \url{https://zenodo.org/record/2541961}), we conclude that the issue simply appears to be inherent to the inductive DAVE4VM estimate. It may arise from the DAVE4VM velocity inversion and/or from our method of extracting the non-inductive contribution from this data, combined with the loss of information in spatial rebinning and lowering the cadence. One particular issue in the inductive DAVE4VM estimate is that even though the raw DAVE4VM electric field fulfills the property $\mathbf{E} \cdot \mathbf{B} = 0$ of the ideal Ohm's law exactly, the inductive DAVE4VM electric field does not.

Finally, we have also plotted a modified version of the PDFI estimate, the PFI (PTD-FLCT-Ideal) estimate to Figures \ref{F-Em_fluxes_div_ED_PDFI_PFI_and_P} and \ref{F-HR_fluxes_div_ED_PDFI_PFI_and_P} (purple curves), which lacks the contribution from Dopplergrams. We can see that this estimate has similar trends in energy and helicity injections as the PDFI estimate until Feb 16 $\sim$00:00, after which the PFI method continues to produce mostly positive injection rates, whereas the PDFI injection rate becomes negative. This comparison reveals the Dopplergram contribution to be the main cause of the negative helicity injection rate after Feb 16 00:00. The Dopplergram contribution is also a likely reason for the differences between the DAVE4VM-based and PDFI helicity injection estimates (Figure \ref{F-HR_fluxes_DAVE4VM_full_res}), as the former do not include the Dopplergram data.

The significant difference in the sign of the helicity injection after Feb 16 00:00 between PDFI and DAVE4VM-based estimates provides another strong distinguishing test for the methods. If the strong negative helicity injection rate of the PDFI method is realistic, this would directly imply that the lack of Dopplergram contribution in DAVE4VM is a serious issue, and the method should be updated to include Dopplergrams \citep[see][for work towards this direction]{Schuck2012}. If the negative injection rate is unphysical, this on the other hand, would indicate an overemphasis or problematic formulation of the Dopplergram contribution in the PDFI method. The Dopplergram contribution is likely overemphasized in the emergence-driven ANMHD test used for validating and optimizing the PDFI method (\citealp{Kazachenko2014}, see also Introduction, Section \ref{S-Introduction}). Moreover, \citet{Liu2014} reported that in a set of 28 active regions all had dominant helicity injection arising from the shear component (i.e. the horizontal component of the DAVE4VM velocity), not the vertical emergence, which we also reproduce for AR 11158. However, this result is derived using the raw DAVE4VM method, and without the Dopplergram contribution. The Dopplergram contribution in the PDFI method might be also contaminated by the horizontal penumbral Evershed flows. As these flows are thought to flow mostly parallel to the magnetic field \citep[e.g.][]{Borrero2011}, they should produce no electric field contribution in ideal Ohm's law, but they still can produce non-zero contribution to the Dopplergram electric field in the PDFI method. The contamination  appears when the AR 11158 moves far from the disk center (central meridian passage on Feb 14 $\sim$02:00 UT) where the Dopplergram velocity $\vec{V}_{LOS}$ becomes increasingly horizontal. Consequently, the Evershed flow contamination presents a possible reason for the strong negative helicity injection rate in the PDFI estimates Feb 16 00:00 UT onward. An alternative version of the PDFI method which attempts to remove the Evershed flow contamination in the Dopplergram electric field contribution has been developed \citep{Fisher2019}, and the signal of the negative helicity injection rate has indeed disappeared in the preliminary tests of this method. However, the alternative method performs worse than the original one in the synthetic ANMHD tests, and produces also problematic artifacts when applied to observational data. Thus, the approach still requires further development and testing. 

All in all, further tests with real and synthetic data are required for both PDFI and DAVE4VM methods to better understand and constrain the issues of the methods discussed above as well as the mutual discrepancies such as the opposite signs of the helicity injection rates. 

\subsection{General discussion}
	\label{S-gen_discuss}
	
To the best of our knowledge, the effect of cadence on the photospheric total energy and helicity injections has not been studied before using real observational input. However, there are certain simulations studies relevant for our work. For example, \citet{Leake2017} studied how well the SDO/HMI cadence of 12 minutes is capable of capturing an idealized flux emergence process in a coronal simulation driven by photospheric data. They varied the effective cadence by changing the emergence rate and compared their data-driven simulation to a ground-truth run of the emergence process, in which the whole emergence from the upper convection zone to the corona was captured. Most of their simulations (7 out of 9) followed a clear trend where increasing the effective cadence produced larger total and free energy budgets in the simulation and producing smaller errors w.r.t. the ground-truth run. This is similar to our results that show increasing or roughly constant energy injections with increasing cadence. 

Two of 9 simulations of \citet{Leake2017} produced exceptions to the trend above resulting in significant overestimation of the free and also the total energy budgets in the simulations (J. Leake, 2019, private communication), when compared to the ground-truth run. They attributed this result to arise from the undersampling of very rapidly evolving photospheric magnetic features ($V_h \sim 20$ km s$^{-1}$ in one of these simulations), which introduced spurious currents to the coronal volume, thus increasing the free and total energy budgets. We mostly did not find any dramatic increases in our total energy injection estimates (except for the 50\% increase between the 2.25- and 11.25-minute raw DAVE4VM estimates, which we however, interpreted as a method-related effect, see Section \ref{S-role_of_nind_disc}). This is not surprise as we did not detect nearly as high horizontal velocities as in the problematic simulations of \citet{Leake2017}, our velocity estimates being typically $<$1 km s$^{-1}$. However, one should note that direct comparison between our results and the findings of \citet{Leake2017} is problematic due to the vast differences in the active regions settings (idealized emergence simulations vs. real observations) as well as the fact that the photospheric injections cannot be compared directly to the coronal budgets. The latter arises both from the fact that injected photospheric energy and helicity may be ejected from the computational domain as well as from the limitations set by the mathematical formulation of the photospheric boundary condition \citep[see e.g.][for further discussion]{Kazachenko2015,Pomoell2019}.

Another simulation study relevant to our work was done by \citet{Weinzierl2016}, who ran a series of global, data-driven magnetofrictional simulations for a reasonably active corona. They used synoptic $B_r$ magnetograms from Air Force Data Assimilative Photospheric Flux Transport (ADAPT) model as input, estimated the inductive electric field consistent with $B_r$ evolution and the non-inductive electric field consistent with the large-scale velocity field related to the solar differential rotation, and drove the simulation using this total electric field. They varied the cadence of the input ADAPT maps between 2 and 48 hours, and compared the effect of driving cadence to the output of the simulations. They reported reasonably consistent results for the cadences of 2 and 6 hours, lower cadences producing notable differences between the simulations. We found the PDFI injection estimates to be consistent within the uncertainties in a cadence range from 2.25 minutes to 2 hours. Our 2-hour limit is smaller but still of similar order of magnitude as the 6-hour limit of \citet{Weinzierl2016}. 

The consistency of our results in a certain cadence range, and the similar findings made by \citet{Weinzierl2016}, have interesting implications on the relevant time scales. The fact that our PDFI energy and helicity injections are consistent (within uncertainties) in the range from 2 minutes to 2 hours imply that the photospheric processes (e.g. flux emergence, cancellation, shearing motions) acting on shorter time scales than that of 2 hours contribute very little to our estimates. This may arise from the actual small role of these processes, or alternatively from the fact that the detection of these processes is below the sensitivity of the SDO/HMI input data and our implementation of the PDFI electric field inversion scheme. One issue that may degrade the sensitivity of the PDFI method is the fact that its perfectly inductive electric field reproduces all magnetogram noise in $\partial \vec{B}/\partial t$ in Faraday's law, unlike the raw DAVE4VM electric field that has been designed to reproduce the $\partial \vec{B}/\partial t$ only in a least squares sense (which, however, produced other issues detailed in Section \ref{S-role_of_nind_disc}). The consistency in the range from 2 minutes to 2 hours implies also that the novel high-cadence 2.25-minute SDO/HMI data brings little new features to the PDFI injection estimates when compared against the nominal 12-minute cadence, which on our study is represented by the mock nominal cadence of 11.25 minutes. Further studies are required to ascertain whether this consistency over various cadences can be reproduced also in the context of data-driven simulations similarly to \citet{Weinzierl2016}. 

\section{Summary and conclusions}
	\label{S-sum_and_concs}
	
In this article we performed an extensive study on the effect of input data cadence on the total photospheric energy and helicity injections, derived from the electric field inversion results of three state-of-the-art methods, the PDFI method, and two methods based on the DAVE4VM velocity inversion approach. The first, the raw DAVE4VM method, follows previous works \citep[e.g.][]{Liu2012} and derives the electric field directly from the output velocity using ideal Ohm's law, whereas the second, the novel inductive DAVE4VM estimate, includes only the horizontal curl-free (non-inductive) part of the raw DAVE4VM field, combined with the inductive field component from the PDFI electric field. 

We produced 6 input magnetogram and Dopplergram data series for eruptive active region NOAA AR 11158 by sampling the novel 2.25-minute high-cadence SDO/HMI data \citep{Sun2017} with cadence varying from 2.25 minutes to 24 hours. We also created one dataset from the nominal 12-minute SDO/HMI data \citep{Hoeksema2014}, as well as additional 4 data series of cadences 2--24 hours where the data was rebinned by a factor of 15 to constrain the effect of undersampling the typical photospheric motions at these low cadences. We optimized the FLCT and DAVE4VM optical flow velocity inversion for each cadence and spatial resolution, inverted the electric field using our three methods and estimated the respective energy and helicity injections. We also derived error bars for the energy and helicity injections using a Monte Carlo approach and the noise characteristics of the input magnetograms. 

The errors arising from the magnetogram noise in the total time-integrated injection estimates were often very small ($<$1\%), in contrast to the previously reported values ($\sim$10\%), which emphasizes the larger uncertainties arising from the inversion methods and data processing. All injection estimates also produced systematic responses when the input magnetograms were manually perturbed with noise, which highlights the need for further tests on the noise response of the methods in controlled environments and using larger number of Monte Carlo statistics than used in this work. Moreover, care should be taken when comparing the injection estimates from data products with variable noise characteristics.

Our three electric field inversion methods produced consistent energy injection estimates at the time of the strongest (X-class) flare of NOAA AR 11158 for the nominal 12-minute HMI cadence \citep[similarly to][]{Kazachenko2015}, but for helicity injections the methods differed significantly, slightly at the time of the flare, and increasingly after the flare. The two DAVE4VM-based methods and PDFI method even gave opposite signs for the helicity injection rate after the flare, and we showed that this difference arises from the Dopplergram contribution in the PDFI method, a contribution not included in the DAVE4VM estimates. This unresolved discrepancy highlights the need for further synthetic tests for both the PDFI and DAVE4VM methods in more realistic settings than offered by the previously employed ANMHD-based tests.

The injection estimates based on the PDFI electric field inversion method turned out to be the most robust against variable cadence, as the cadences from 2.25 minutes to 2.025 hours produced all energy and helicity injections consistent within the uncertainties at the time of the strongest X-class flare of NOAA AR 11158. This result suggests that the PDFI method can be used flexibly with data products of variable cadence from other instruments beside SDO/HMI \citep[see][for a review]{Lagg2017}, but also with the SDO/HMI data in the case of data gaps, or when lowering the cadence in pursuit of saving computational resources. This result also implies that the photospheric processes (e.g. flux emergence, cancellation and shearing motions) acting on time scales below 2 hours contribute little to the energy and helicity injections, or alternatively that the detection of these contributions is below the sensitivity of the SDO/HMI instrument and/or our implementation of the PDFI electric field inversion scheme. Furthermore, this result implies that the novel high-cadence 2.25-minute SDO/HMI data brings little new features to the total injections when compared against the nominal 12-minute cadence (in NOAA AR 11158).

The injection estimates based on the raw DAVE4VM estimates turned out to be very sensitive to cadence: already the 2.25- and 11.25-minute cases differed significantly, particularly in the total energy injection where the 2.25-minute input produced roughly 50\% larger injection. For the lowest cadences ($\geq$2 h) the raw injection estimates lost the injection signal altogether producing zero or negligibly small injections.

The inductive DAVE4VM estimate turned out to remedy most of the large difference between the 2.25- and 11.25-minute raw DAVE4VM energy injection estimates proving that the difference arises mostly from the poor and variable inductivity of the raw DAVE4VM electric field estimate. Based on this finding, we recommend correcting for the inductivity when deriving the evolutionary energy and helicity injection estimates from the DAVE4VM electric fields, a correction not included in the previous studies employing the method. The loss of the raw DAVE4VM injection signal for the lowest cadences ($\geq$2 h) was also visible in the inductive DAVE4VM estimate, for which the injections were clearly non-zero, but arose is most cases predominantly from the inductive component of the electric field, and still clearly underestimated the high-cadence estimates.

The correction of the severe undersampling for the lowest cadences ($\geq$2 h) by rebinning the input data by a factor 15, clearly improved only the raw DAVE4VM injection estimates, rebinned input data producing about half of the 11.25-minute raw DAVE4VM reference estimates at the time of the X-class flare in NOAA AR 11158. On the other hand, rebinned input only degraded the inductive DAVE4VM and PDFI estimates producing spurious negative helicity injection results for the former and significant (up to 90\%) loss of helicity signal  for the latter.

We mostly did not find dramatic increases in the total energy injections for any the cadences or electric field inversion methods, but instead found the energy injections to decrease as a function of decreasing cadence. As this holds also when our input data clearly undersamples typical photospheric motions, we can conclude that our results show no evidence for any spurious energy injections arising from the undersampling of the photospheric motions, such as the ones detected by \citet{Leake2017}.

Though most of the very low-cadence ($\geq$6 h) estimates considered in this study clearly underestimated the injections, we still find that estimates, such as the low-cadence full resolution PDFI estimates, and 15-times rebinned DAVE4VM estimates, are capable of approximating temporal trends of the high-cadence reference estimates reasonably well, and even giving crude (under)estimates of the absolute values. For example, the 12-hour full resolution PDFI estimate produced 50\% and 74\% of the 11.25-minute reference energy and helicity injections at the time of the X-class flare. This implies that even this low cadences are capable of capturing some of the large scale evolution in the NOAA AR 11158 (possibly related to the fact that significant changes in the magnetic field structure of active regions occur over a time scale of several hours, \citealp{Fisher1998,Metcalf1994,Pevtsov1994}).

Movies of all the magnetogram, Dopplergram, optical flow, electric field as well energy and helicity flux maps for all of our data series can be found from \url{https://zenodo.org/record/2541961}.

\begin{acks}
All of the data download, magnetogram/Dopplergram processing steps, FLCT velocity inversion, and PDFI electric-field inversion are handled using the ELECTRIC field Inversion Toolkit, ELECTRICIT software \citep{Lumme2017}. When it comes to the Dopplergram calibration, FLCT velocity inversion and PDFI electric-field inversion, the toolkit acts as a wrapper for a Fortran Dopplergram calibration library by B. T. Welsch, the FLCT code v1.06 (\url{http://solarmuri.ssl.berkeley.edu/~fisher/public/software/FLCT/}) and the PDFI\_SS software (\url{http://cgem.ssl.berkeley.edu/cgi-bin/cgem/PDFI_SS}). We used the latest public version of the DAVE4VM code (\url{https://ccmc.gsfc.nasa.gov/lwsrepository/index.php}), which was executed outside ELECTRICIT using Interactive Data Language (IDL) scripts based on the work by \citet{Schuck2008}. This research has made use of SunPy, an open-source and free community-developed solar data analysis package written in Python \citep{Mumford2015}. We thank the HMI team for providing us with the vector magnetic-field and Dopplergram data.

E. Lumme acknowledges the doctoral program in particle physics and universe sciences (PAPU) of the University of Helsinki, and Emil Aaltonen Foundation for financial support. E. Lumme, J. Pomoell and E.K.J. Kilpua acknowledge the SolMAG project (ERC-COG 724391) funded by the European Research Council (ERC) in the framework of the Horizon 2020 Research and Innovation Programme, and the Finnish Centre of Excellence in Research of Sustainable Space (Academy of Finland grant number 1312390). M.D. Kazachenko acknowledges National Science Foundation, SHINE, AGS 1622495 award. M.D. Kazachenko, G.H. Fisher, and B.T. Welsch acknowledge the Coronal Global Evolutionary Model (CGEM) award by the joint NSF-National Space Weather Program / NASA-LWS Strategic Capability program under award NSF AGS 1321474. This award supported also the development of the Doppler and PDFI\_SS packages. B.T. Welsch acknowledges support from the NASA-LWS Targeted Research \& Technology under award 80NSSC19K0072. 

This is a pre-print of an article published in Solar Physics. The final authenticated version is available online at: \url{https://doi.org/10.1007/s11207-019-1475-x}.
\end{acks}

\appendix

\section{Technical details related to the velocity and electric field inversions}
	\label{S-tech_detail_appendix}
This section discusses the technical details related to the temporal and spatial discretization of the velocity and electric field inversion, as well as the updates made to the PDFI method after the description of \citet{Kazachenko2014} and how we employ the updated version.

\subsection{Time discretization in the velocity and electric field inversions}
	\label{S-time_dscr_v_and_einv}
	
When specifying the input data for the FLCT and DAVE4VM velocity inversions we employ a central difference scheme where magnetograms at times $t$ and $t \pm \Delta t$ are used to estimate the velocity at time $t$ so that the time derivatives $\partial B_z/\partial t$ and $\partial \vec{B}/\partial t$ are both estimated using from $\vec{B}(t\pm \Delta t)$ using a central difference scheme, and the spatial derivatives $\partial_l B_m$ ($l \in \{x,y\}$ and $m \in \{x,y,z\}$) required for DAVE4VM are computed from the central magnetogram $\vec{B}(t)$ at time $t$ (using 5-point optimized derivatives, \citealp{Jahne2004,Schuck2008,Liu2012}). The central difference scheme that employs three magnetograms at $t$ and $t \pm \Delta t$ is preferred because it minimizes the effect of correlated noise between the estimates of temporal and spatial derivatives \citep{Welsch2007,Fermuller2001}, and improves also the final electric field estimates through the ``staggered-in-time'' approach as explained below.

As opposed to the velocity inversion, we employ a different time discretization in the electric field inversion, which we refer to as ``staggered-in-time'' approach \citep[see also][]{Fisher2019}. Essentially, we use consecutive magnetograms at time $t$ and $t + \Delta t$ to estimate $\partial \vec{B}/\partial t(t+1/2\Delta t)$ at time $t + 1/2\Delta t$ as follows:
\begin{equation}
\frac{\partial \vec{B}}{\partial t}(t+1/2\Delta t) = \frac{\vec{B}\left(t+\Delta t\right) - \vec{B}(t)}{\Delta t},
\end{equation}
where $\Delta t$ is the cadence of the input data series. Since the time derivative is estimated at time $t+1/2\Delta t$, the electric field inversion must be done at the same temporal position, which requires (linear) interpolation (i.e. averaging of two consecutive frames) of all required quantities ($\vec{B}$, $\vec{V}_{LOS}$, $\vec{V}_h^{FLCT}$, $\vec{V}_{DAVE4VM}$) to this same position before doing the inversion. This approach is beneficial as it offers simplicity for the physical interpretation of the electric field: a perfectly inductive electric field $\vec{E}(t+1/2\Delta t)$ drives the magnetic field exactly from observed value $\vec{B}(t)$ to the next $\vec{B}(t+\Delta t)$, which does not happen for a standard central difference discretization. Moreover, the approach partly mitigates some of the issues caused by the noise, and the effect of other artefacts such as the spurious velocity spikes of the FLCT inversion (Appendix \ref{S-mask_in_v_and_einv}). 

On the other hand, this choice of discretization, combined with the central difference scheme used in the velocity inversion means that $\vec{E}(t+1/2\Delta t)$ at time $t+1/2\Delta t$ depends on the input magnetograms ranging from $t-\Delta t, t+2\Delta t$ where the most distant magnetograms are included via the central difference scheme of the optical flow estimates. Therefore, when considering electric field of cadence $\Delta t$ some of the input data is actually collected from an interval of length $3\Delta t$. 

\subsection{Masking in the velocity and electric field inversion}
	\label{S-mask_in_v_and_einv}

In order to remove the noise-dominated pixels from the output we employ a noise threshold of $|\vec{B}| = 300$ Mx cm$^{-2}$ both in the DAVE4VM and FLCT velocity inversions. However, as explained by \citet{Lumme2017} the input to DAVE4VM cannot be masked, and therefore the masking of the DAVE4VM estimates is performed post facto when computing the electric field (see below). For FLCT we employ the masking functionality implemented into the method itself by setting the \verb+thr+ parameter to 300 Mx cm$^{-2}$, in which case pixels where $|B_z(t_1)+B_z(t_2)|/2 < $ \verb+thr+ are excluded from the analysis and the output velocity is set to zero \citep{Fisher2008}. This choice produces a stronger masking condition for the FLCT inversion than the $|\vec{B}| = 300$ Mx cm$^{-2}$ threshold. We find this, however, beneficial due to the sensitivity of the FLCT to noise, resulting for example in a larger number of spurious velocity spikes (where velocity gets unreasonably high $|\vec{V}_h| > 2$ km s$^{-1}$) with the lower choices of \verb+thr+ value.
	
When it comes to masking the noise-dominated pixels of the input data in the electric field inversion, we follow mostly the approach of \citet{Kazachenko2015} modified to the updated spatial (Appendix \ref{S-app_mods_to_PDFI}) and temporal (Section \ref{S-time_dscr_v_and_einv}) discretization of the inversion procedures \citep[see also][]{Fisher2019}. For masking the magnetic field input data we use a fixed masking threshold of $|\vec{B}| = 300$ Mx cm$^{-2}$, which is employed consistently with the staggered-in-time temporal discretization. This means that the magnetic field input $\vec{B}(t+1/2\Delta t)$, $\partial \vec{B}/\partial t(t+1/2\Delta t)$ for the electric field inversion at $t+1/2\Delta t$ is masked in pixels where either of the magnetograms participating in the inversion $\vec{B}(t)$, $\vec{B}(t+\Delta t)$ is below the threshold. Since in the PDFI method the input magnetic field data is interpolated to staggered grid positions (following the spatial discretization presented in Appendix \ref{S-app_mods_to_PDFI}) the mask is interpolated consistently so that the fractional values that arise from the interpolation of the $\{0,1\}$ mask are rounded up in the process. Except for the independent masking of the FLCT method (see Section \ref{S-vel_inversion}) other input data are not masked in any way, since the masking of the magnetic field data handles this implicitly in the application of the ideal Ohm's law. 

\subsection{Use of the PDFI\_SS software in this study}
	\label{S-app_mods_to_PDFI}

We use the public version of the PDFI\_SS (``PDFI\_Spherical\_Staggered'') software downloaded from \url{http://cgem.ssl.berkeley.edu/cgi-bin/cgem/PDFI_SS/index} on Aug 30, 2018 16:43 GMT. The software has several differences to the PDFI version described in the latest publication by \citet{Kazachenko2014}. The differences are presented in detail by \citet{Fisher2019}, but we also briefly summarize them below:	
\begin{itemize}
\item Instead of Cartesian geometry spherical coordinates are used consistently throughout the electric field inversion.
\item All vectors are given in a local spherical basis $(\vec{e}_r,\vec{e}_{\theta},\vec{e}_{\phi})$ corresponding to spherical coordinates on the surface of the Sun $(R_{\hbox{$\odot$}},\theta,\phi)$. We employ this coordinate system so that the center of the input data patch is assumed to be at $(\theta,\phi) = (90^{\circ},0^{\circ})$.
\item All spatial derivatives are computed in the spherical coordinate system described above \citep[see][]{Kazachenko2014,Fisher2019}.
\item Before computing the electric field the input data (i.e. the magnetic field and plasma velocity estimates) are interpolated to a staggered grid \citep{Yee1966,Fisher2019} in two dimensions. As a short summary, the grid is defined so that the input data are specified in the cell corners, the output horizontal electric field components $(E_{\theta},E_{\phi})$ are given at cell edges, staggered w.r.t. cell centers, and the output radial component $E_r$ is specified at the cell corners. When the software is used to compute the magnetic vector potential the output components $(A_r,A_{\theta},A_{\phi})$ are defined at the same grid positions as the electric field.
\item All spatial derivatives are computed consistently with the discretization of the Poisson equations in the inversion, i.e. the fact that the 5-point stencil assumes 1st order spatial derivatives to be defined at half-grid points is taken into account \citep{Kazachenko2014,Lumme2017,Fisher2019}. This ensures that our electric field estimate is exactly inductive with respect to the magnetic field input interpolated into the staggered grid positions (except for small numerical errors discussed below), where the $B_r$ component is interpolated to cell centers. 
\end{itemize}

Since we wish to employ Cartesian instead of spherical geometry in our inversion (see Section \ref{S_einv_PDFI}), the default inversion scheme in PDFI\_SS must be modified. Following the documentation in the PDFI\_SS library we define a ``pseudo-Cartesian'' coordinate system on the surface of a very large sphere with radius $R' \gg R_{\hbox{$\odot$}}$, where the latitudinal extent of our patch is fixed to be very small, $10^{-4}$ radians:
\begin{equation}
N_{\theta}\Delta \theta' = 10^{-4} \ \textrm{rad} \approx 0.0057^{\circ}.
\end{equation} 
Here $N_{\theta}$ is the number of cells in our patch in latitudinal $\theta$ direction (for our $547 \times 527$ patch $N_{\theta} = 527 - 1 = 526$), and $\Delta \theta'$ is the grid spacing in the pseudo-Cartesian system. Now we choose the radius of this large sphere so that the physical size of our original patch as well as the pixel size ($\Delta y = R_{\hbox{$\odot$}}\Delta \theta = R_{\hbox{$\odot$}}\Delta \phi = \Delta x$) remain the same (for our patch with $\Delta \theta = \Delta \phi = 0.03^{\circ}$, $\Delta y = \Delta x \approx 364$ km). This gives relation for the radius of the large sphere $R'$:
\begin{eqnarray}
N_{\theta}\Delta y' &=& N_{\theta}\Delta y \\
N_{\theta} R' \Delta \theta' &=& N_{\theta} R_{\hbox{$\odot$}} \Delta \theta \\
R' &=& \left(\frac{N_{\theta} \Delta \theta}{N_{\theta} \Delta \theta'}\right)R_{\hbox{$\odot$}} \\
R' & \approx & 2750R_{\hbox{$\odot$}}
\end{eqnarray}
Using the pseudo-Cartesian coordinate system $(R',\theta',\phi')$ when calling the PDFI\_SS software we preserve all the properties of the original input data patch -- including the grid spacing as well as the magnitudes and LOS vector directions of the Dopplergram velocities -- but the spherical corrections in the spatial derivatives become vanishingly small and all the computations are approximately Cartesian. Finally, the output is transformed as: $(E_x,E_y,E_z) = (E_{\phi},-E_{\theta},E_r)$ following the definition of a local Cartesian basis from \citet{Lumme2017}. 

The maximum error introduced into the inductivity of the output electric field by the pseudo-Cartesian approximation is:
\begin{equation}
\label{Eq-ind_error}
\frac{\max |(\nabla \times \vec{E})_z - (-\partial B_z/\partial t)|}{\left<|\partial B_z/\partial t|\right>} \lesssim 10^{-8},
\end{equation}
when compared to the average magnitude of $|\partial B_z/\partial t|$ (excluding the masked noise-dominated pixels, $|\vec{B}| < 300$ Mx cm$^{-2}$). This error is much smaller than the error introduced by the FISHPACK Poisson solver \citep{Swarztrauber1975} in the PDFI software, which removes the means of the source terms of the Poisson equations used in solving the inductive electric field using Poloidal-Toroidal Decomposition \citep[see][]{Kazachenko2014,Lumme2017}. This produces an additional error of $\sim 10^{-2}$ to Eq. \ref{Eq-ind_error} above. Although it is possible to remove this additional error altogether by using different boundary conditions for the solutions of the Poisson equations \citep{Fisher2019} or alternatively post facto electric field corrections \citep[see][]{Fisher2010,Lumme2017}, we decided not to employ either of these due to the fact that the error produced by this issue remained small for the well-isolated active region NOAA 11158 used in this study.

\section{Optimization of the windowing parameters of FLCT and DAVE4VM}
	\label{S-opt_res_FLCT_and_D4VM}
	
As discussed in Section \ref{S-vel_inversion} we optimize the windowing parameters of FLCT ($\sigma_{FLCT}$) and DAVE4VM (square top hat side length) for each of our data series of variable cadence and spatial resolution so that the output velocities fulfill the advection equation (Eq. \ref{Eq-adv_eq_FLCT}) and the normal component of the induction equation (Eq. \ref{Eq-n_induc_eq}), respectively, as well as possible. More specifically, we measure the success in the reproduction of these equations by the following metrics \citep[also used by][]{Schuck2008}: 
\begin{eqnarray}
\label{Eq-opt_perf_mtr_1}
& \rho \ \textrm{slope in the fit:} \ T = \rho X + \alpha \\
\label{Eq-opt_perf_mtr_2}
& C \ \textrm{Pearson correlation:} \ C(T,X) \\
\label{Eq-opt_perf_mtr_3}
& S \ \textrm{Spearman rank order correlation:} \ S(T,X)
\end{eqnarray}
where $T = \partial B_z/\partial t$, the time derivative term in the advection/induction eq., and $X$ is the spatial derivative term, $\vec{V}_h \cdot \nabla_h B_z$ in the advection equation (Eq. \ref{Eq-adv_eq_FLCT}) and $\nabla_h \cdot (B_z\vec{V}_h - V_z \vec{B}_h)$ in the induction equation (Eq. \ref{Eq-n_induc_eq}). Since optimal reproduction of the respective equations yields $T = -X$, the optimal metrics are thus $\rho = C = S = -1$. Though we consider all of these metrics in our optimization procedures, we nonetheless aim to find a single metric for FLCT and DAVE4VM optimizations each so that the metric works robustly across all cadences and spatial resolutions. 

\begin{figure}[htb]   
    \centerline{\includegraphics[width=\textwidth, trim = 0.0cm 0.0cm 0cm 0.0cm]{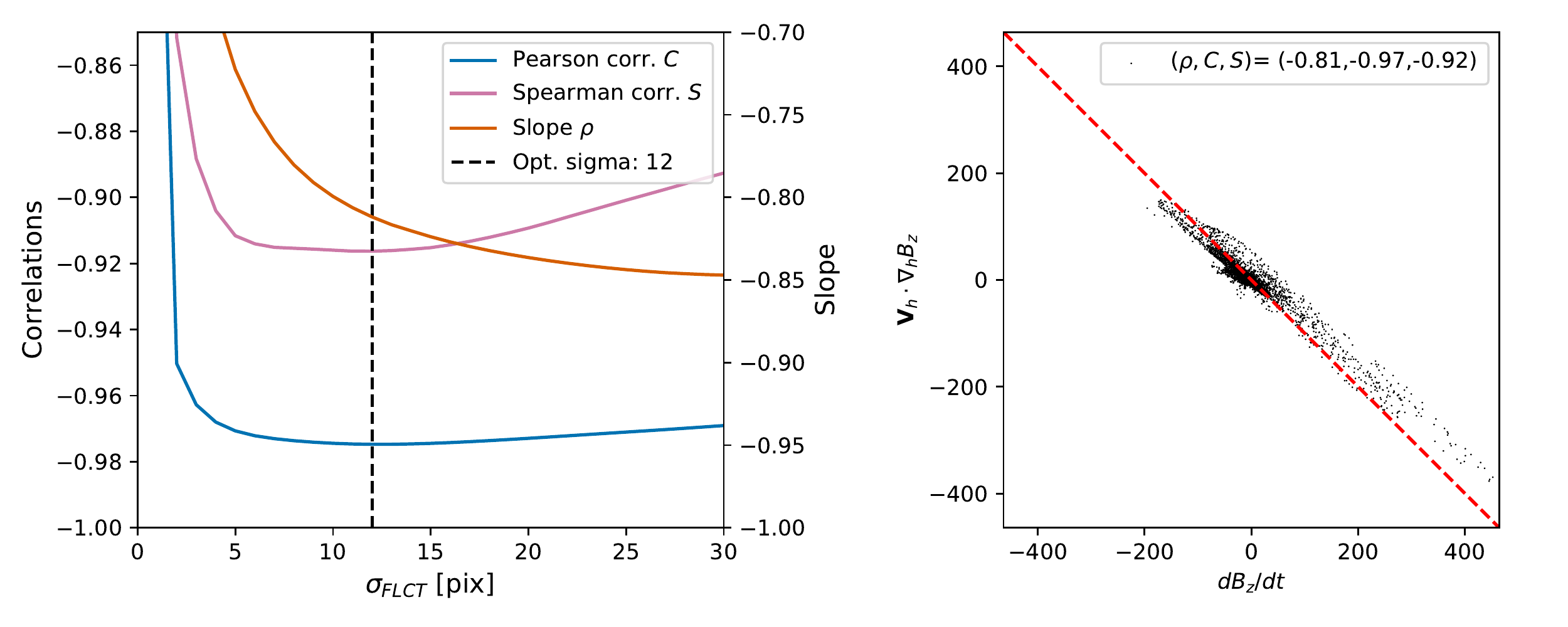}}  
   \caption{Optimization of $\sigma_{FLCT}$ using the synthetic magnetogram data from ANMHD simulation. Left panel shows the optimization curves for the metrics slope, Pearson and Spearman correlation, where the optimal $\sigma_{FLCT} = 12$ pixel at the minimum of Pearson correlation has been indicated by the vertical dashed line. Right panel shows the scatter plot between the two terms of the advection equation for the optimal $\sigma_{FLCT} = 12$ pixels. The red dashed line shows the $y=-x$ line $\vec{V}_h \cdot \nabla_h B_z = -dB_z/dt$, to which the scatter plot points should fall in the case of perfect reproduction of the advection equation.} 
   \label{F-ANMHD_results_FLCT}
\end{figure}

For testing our optimization approach we employ synthetic magnetogram data from the ANMHD simulation (\citealp{Abbett2004,Welsch2007,Schuck2008}, see also Introduction, Section \ref{S-Introduction}), which has been used to test velocity inversion routines. Since our optimization algorithm for the DAVE4VM window size is essentially the same as the one used by \citet{Schuck2008}, we were able to reproduce their optimization curves for metrics $\rho$, $C$ and $S$ above (see \citealp{Schuck2008}, Figure 1, upper right panel). $\sigma_{FLCT}$ on the other hand has been previously optimized to give the best consistency between the output FLCT velocity and the horizontal components of the known ANMHD velocity field \citep{Welsch2007,Kazachenko2014}, which is fundamentally different from our approach. When applied to ANMHD data our optimization algorithm gives the optimal value $\sigma_{FLCT} = 12$ pixels from the Pearson and Spearman correlation metrics, whereas the slope metric reaches no proper minimum in the optimization interval (see Figure \ref{F-ANMHD_results_FLCT}, left panel). The optimal value of 12 pixels is close to the optimal value of 15 pixels by \citet{Welsch2007} and \citet{Kazachenko2014}, and the difference is acceptable considering the flatness of the $C(\sigma_{FLCT})$ and $S(\sigma_{FLCT})$ curves near the values $\sigma_{FLCT} \in [10,15]$. Based on this test, the correlation metrics appear to yield the most unique optimization for $\sigma_{FLCT}$, and in the further optimizations discussed below, we found that the Pearson correlation gives the most robust results. Thus, hereafter we define the optimal $\sigma_{FLCT}$ to be at the minimum of the Pearson correlation metric (Eq. \ref{Eq-opt_perf_mtr_2}) of the advection equation. 

\begin{figure}[htb]  
    \centerline{\includegraphics[width=\textwidth, trim = 0.0cm 0.0cm 0cm 0.0cm]{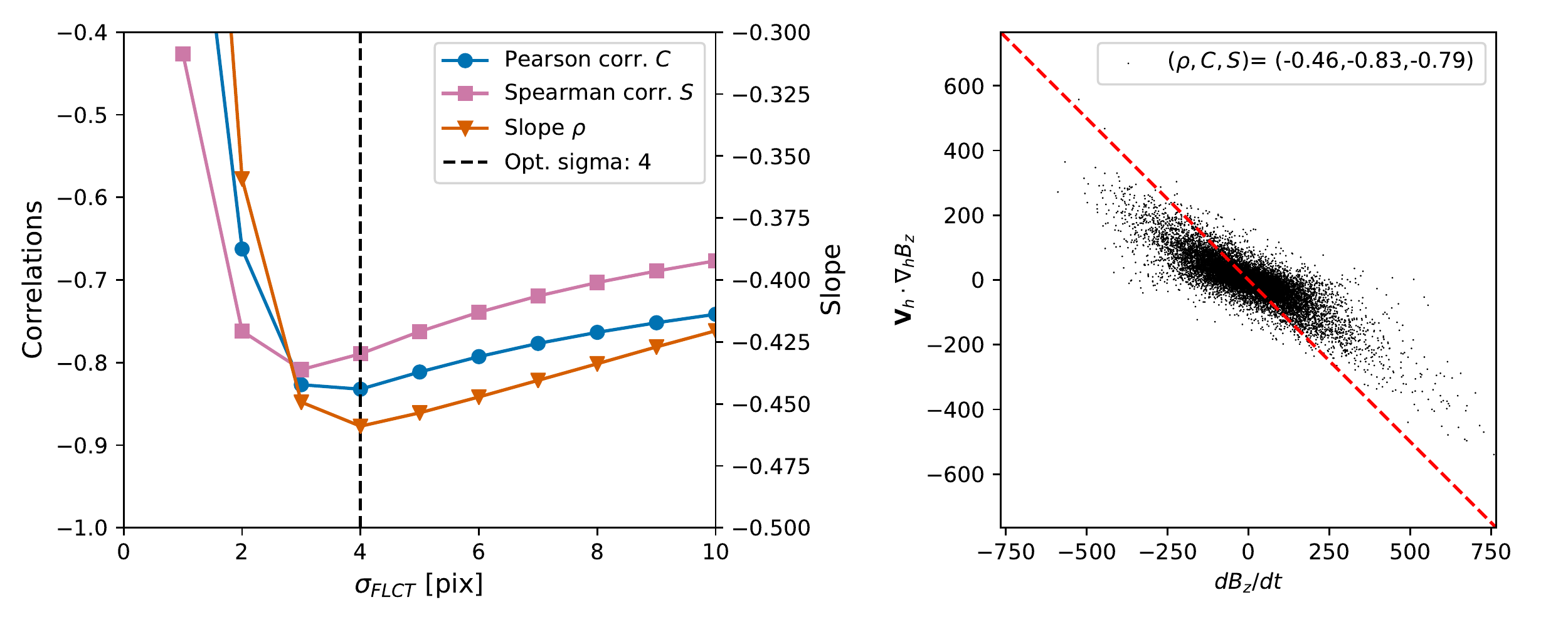}}  
   \caption{Same as Figure \ref{F-ANMHD_results_FLCT}, but now the $\sigma_{FLCT}$ optimization procedure is applied to a representative frame at Feb 14 01:48 in our 12-minute vector magnetogram series.} 
   \label{F-720s_results_FLCT}
\end{figure}

After validating the optimization algorithm against previous ANMHD tests, we moved onto applying the method to real magnetogram observations. We employ a representative frame closest to the central meridian passage of our NOAA AR 11158 series (Feb 14 01:48 UT) for the optimization, which is chosen for its good representativeness of the active region properties (see Section \ref{S-error_bars_disc} for details). As illustrated by Figure \ref{F-720s_results_FLCT} the FLCT optimization for 12-minute data input gives a minimum Pearson correlation at $\sigma_{FLCT} = 4$ pixels (whereas the slope and Spearman correlation metrics yield optimal $\sigma_{FLCT}$ of 4 and 3 pixels, respectively). This is similar to the $\sigma_{FLCT} = 5$ pixels used by \citet{Kazachenko2015}. Unlike the ANHMD test case where all metrics were very good (optimally ${\rho},C,S = \textrm{-}0.81,\textrm{-}0.97,\textrm{-}0.92$, see Figure \ref{F-ANMHD_results_FLCT}, right panel) for real data input the optimal correlations degrade to $\sim$-0.8 and the slope drops to $\textrm{-}0.46$. This is most likely a noise-related effect, since real magnetogram input is clearly more noise-dominated both spatially and temporally than the extremely smooth ANMHD input \citep[see e.g.][for further details on the ANMHD data properties]{Welsch2007}. Table \ref{T-optimization_results} lists the $\sigma_{FLCT}$ optimization results for all cadences and spatial resolutions considered in this study.

\begin{table}[hbt]
\caption{Optimization results for $\sigma_{FLCT}$ and DAVE4VM window size  over all cadences and spatial resolutions considered in this study.
}
\label{T-optimization_results}
\begin{tabular}{c|cccc}     
  \hline                   
 & \multicolumn{2}{c}{Opt. FLCT} & \multicolumn{2}{c}{Opt. DAVEVM}  \\
Dataset & $\sigma_{FLCT}$ & [$^{\rho}C_S$]  & window size & [$^{\rho}S_C$]\\
cadence & [pix] &at opt. $\sigma_{FLCT}$ & [pix] & at opt. w.size \\
\hline
2.25 min & 2 & $^{\textrm{-}0.29}\textrm{-}0.76_{\textrm{-}0.74}$ 
& 19 & $^{\textrm{-}0.26}\textrm{-}0.36_{\textrm{-}0.39}$ \\
11.25 min & 3 & $^{\textrm{-}0.44}\textrm{-}0.81_{\textrm{-}0.79}$ 
& 19 & $^{\textrm{-}0.46}\textrm{-}0.52_{\textrm{-}0.61}$ \\
12 min \tabnote{Different data source, see Sections 2.1 and 2.2 for details.}  & 4 & $^{\textrm{-}0.46}\textrm{-}0.83_{\textrm{-}0.79}$ 
& 19 & $^{\textrm{-}0.52}\textrm{-}0.56_{\textrm{-}0.65}$ \\
2.025 h & 26 & $^{\textrm{-}0.60}\textrm{-}0.59_{\textrm{-}0.58}$
& 29 & $^{\textrm{-}0.36}\textrm{-}0.52_{\textrm{-}0.58}$ \\
- - (rebin 15x) & 30 \tabnote{In full resolution units.} & $^{\textrm{-}0.20}\textrm{-}0.75_{\textrm{-}0.80}$
& 75 $^2$ & $^{\textrm{-}0.55}\textrm{-}0.66_{\textrm{-}0.69}$ \\
6 h & 44 & $^{\textrm{-}0.47}\textrm{-}0.36_{\textrm{-}0.37}$ 
& 45 & $^{\textrm{-}0.17}\textrm{-}0.40_{\textrm{-}0.44}$ \\
- - (rebin 15x) & 45 $^2$ & $^{\textrm{-}0.37}\textrm{-}0.77_{\textrm{-}0.67}$
& 75 $^2$ & $^{\textrm{-}0.51}\textrm{-}0.68_{\textrm{-}0.74}$ \\
12 h & 41 & $^{\textrm{-}0.64}\textrm{-}0.32_{\textrm{-}0.35}$ 
& 39 & $^{\textrm{-}0.18}\textrm{-}0.41_{\textrm{-}0.43}$ \\
- - (rebin 15x) & 45 $^2$ & $^{\textrm{-}0.48}\textrm{-}0.74_{\textrm{-}0.63}$
& 105 $^2$ & $^{\textrm{-}0.52}\textrm{-}0.68_{\textrm{-}0.77}$ \\
24 h & 27 & $^{\textrm{-}0.16}\textrm{-}0.08_{\textrm{-}0.11}$ 
& 39 & $^{\textrm{-}0.12}\textrm{-}0.28_{\textrm{-}0.34}$ \\
- - (rebin 15x) & 30 $^2$ & $^{\textrm{-}0.26}\textrm{-}0.44_{\textrm{-}0.41}$
& 165$^2$ & $^{\textrm{-}0.17}\textrm{-}0.50_{\textrm{-}0.41}$ \\
  \hline
  
\end{tabular}
\end{table} 

\begin{figure}[htb]  
    \centerline{\includegraphics[width=\textwidth, trim = 0.0cm 0.0cm 0cm 0.0cm]{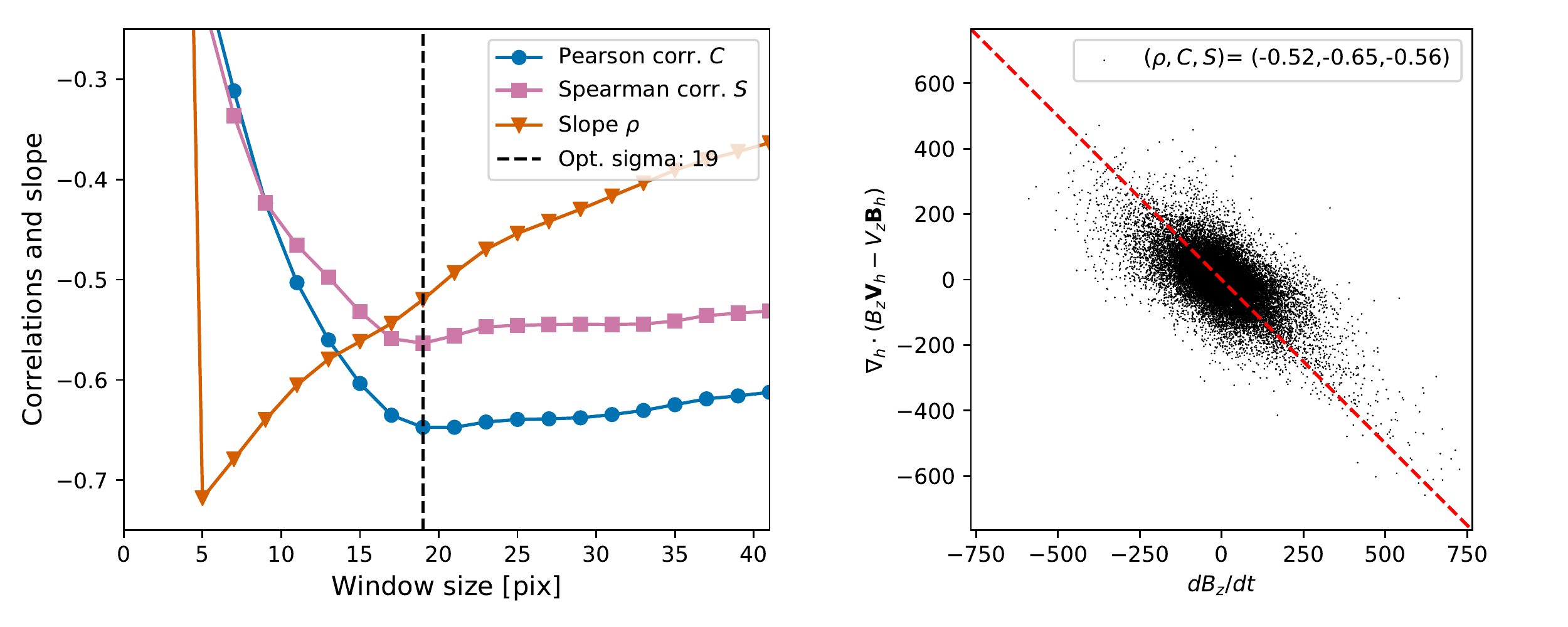}}  
   \caption{Optimization of the DAVE4VM window size using vector magnetogram input from our 12-minute data series. Left panel shows the optimization curves for the slope, Pearson and Spearman correlation metrics with the optimal window size (i.e. the local minimum of Spearman correlation with the smallest window size) indicated by the vertical dashed line. Right panel shows the scatter plot between the two terms of the advection equation for the optimal window size of 19 pixels. The red dashed line shows the $y=-x$ line $\nabla_h \cdot (B_z \vec{V}_h -V_z \vec{B}_h) = -dB_z/dt$, to which the scatter plot points should fall in the case of perfect reproduction of the induction equation.} 
   \label{F-720s_results_DAVE4VM}
\end{figure}

Unlike the $\sigma_{FLCT}$ optimization the DAVE4VM window size case produces clearly more ambiguous results both for ANMHD \citep[see][Figure 1, upper right panel]{Schuck2008} and real observations. As illustrated by the left panel of Figure \ref{F-720s_results_DAVE4VM}, all metrics have a clear minimum for the 12-minute magnetogram input data, but the minima of the correlation and slope metrics are very different ($\sim$19 and $5$ pixels, respectively). Moreover, as already noted by \citet{Schuck2008} the slope metric degrades as a function of the window size, whereas the correlation metrics continue to improve (until saturating at $\sim$19 pixels in our case). Due to the inconsistency between the evolution of the metrics \citet{Schuck2008} suggests that the optimal value is chosen as a balance between the degrading slope metric and improving correlation metrics. 

Based on a survey covering all optimization curves over all cadences and spatial resolutions, we have chosen the optimal DAVE4VM window size to be the one where the Spearman correlation reaches a local minimum with the smallest window size. This approach turned out to be robust over all cadences and spatial resolutions used in this study. More specifically, we find that: (1) a local minimum always existed and it was also often the global minimum, thus emphasizing the uniqueness of the solution, and (2) the optimal window size coincided well with the idea of \citet{Schuck2008} that a balance between the degrading slope and improving correlation metrics should be found. Using this method we obtained an optimal value of 19 pixels for the 12-minute magnetogram input, which is consistent with \citet{Liu2012} and \citet{Lumme2017}.

As shown by Figures \ref{F-720s_results_FLCT} and  \ref{F-720s_results_DAVE4VM}, and further by Table \ref{T-optimization_results} the values of the optimization metrics are in most cases clearly poorer for the DAVE4VM optimization than for $\sigma_{FLCT}$ case. This emphasizes the issue of poor inductiveness of the raw DAVE4VM electric field already discussed in Section \ref{S_einv_DAVE4VM} (as perfect inductivity would mean also perfect metrics $\rho = C = S = -1$). The metrics and inductivity are also clearly worse for real magnetogram input than for the ANMHD data input, e.g. $S = \textrm{-}0.56$ for the 12-minute input, as opposed to optimal $S \sim \textrm{-}0.8$ for the ANMHD input \citep[Figure 1, upper right panel]{Schuck2008}. This indicates that the larger spatial and temporal noise and higher spatial structuring of the real magnetogram data compared to the very smooth ANMHD data is a likely cause for the degration in the inductivity. As shown by the last column of Table \ref{T-optimization_results} the inductivity also changes as a function of cadence, e.g. the optimal Spearman correlation increases from $\textrm{-}0.52$ to $\textrm{-}0.36$ between the mock nominal cadence of 11.25 min and highest cadence of 2.25 min. This issue is further discussed in Section \ref{S-role_of_nind_disc}.

\begin{figure}[htb]  
    \centerline{\includegraphics[width=\textwidth, trim = 0.0cm 0.0cm 0cm 0.0cm]{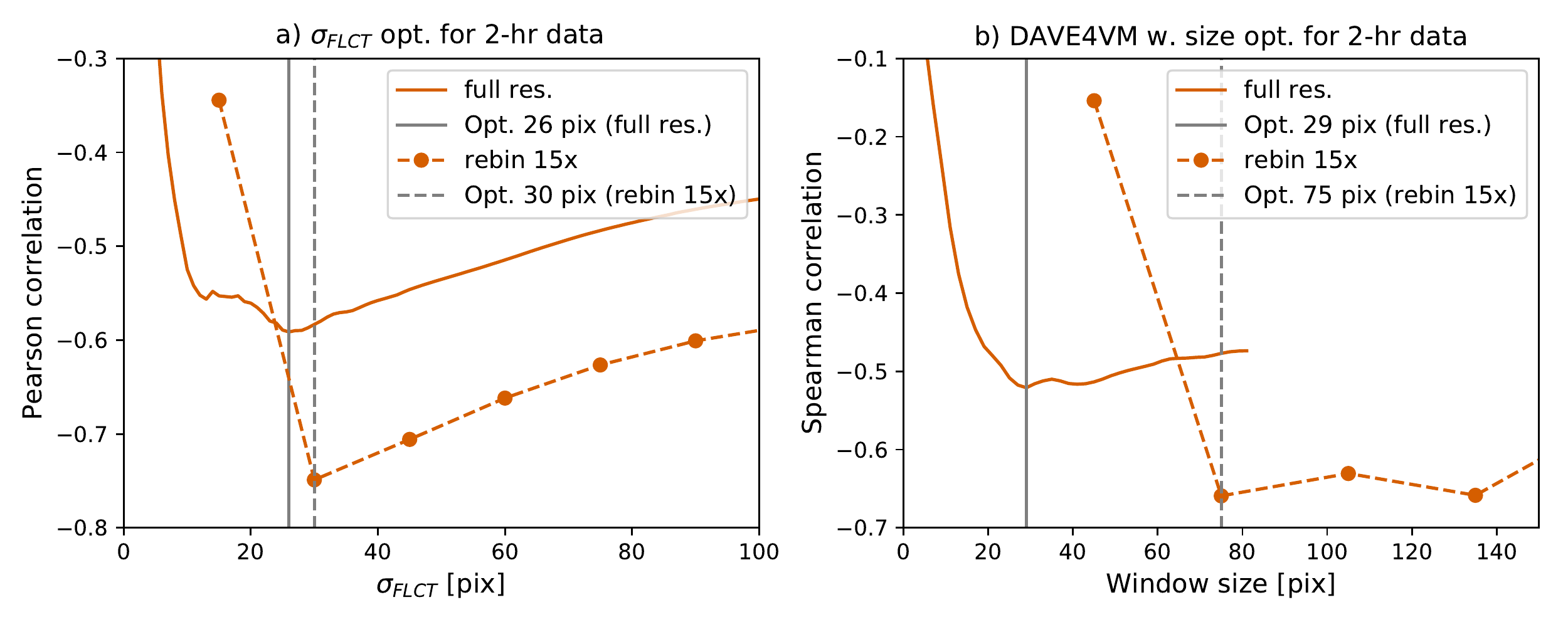}}  
   \caption{Optimization of $\sigma_{FLCT}$ (panel a) and the DAVE4VM window size (panel b) using vector magnetogram input from our 2.025-h data series with full spatial resolution (solid lines) and 15-times rebinned input data (dashed lines). The optimal values for each are indicated by gray vertical lines. Note that $\sigma_{FLCT}$ and DAVE4VM window size in the rebinned case are scaled to the full resolution units.} 
   \label{F-opt_res_FLCT_DAVE4VM_2hr}
\end{figure}

Figure \ref{F-opt_res_FLCT_DAVE4VM_2hr} (solid curves) illustrates the results of the optimization scheme described above for one of the lowest cadences of 2 hours. As indicated by the figure and further by Table \ref{T-optimization_results} most of the metrics become clearly worse at optimal $\sigma_{FLCT}$/window size over the lowest cadences $\geq$2 h. For example, the optimal Pearson correlation in the FLCT optimization drops from $\sim$-0.8 to -0.59 -- -0.08 between the 12-minute and the $\geq $2-hour cadences. This degradation in the quality of the optimization as a function of cadence is not a surprise considering the fact that when using the lowest cadences our central difference approximation for the advection/induction equation compares magnetograms $2\Delta t \in [4,48]$ hours apart. FLCT tracking recovers shifts of tens of pixels over these intervals, and such large shifts make our discretization of the advection/induction equation an extremely poor approximation; optimally the shifts should be of subpixel magnitude \citep[e.g.][]{Welsch2012}. As already discussed in Section \ref{S-sampling_data} we attempt to remove this issue by creating an additional rebinned vector magnetogram time series for the lowest cadences ($\geq$2 h) with 15 times lower spatial resolution. For this dataset the largest FLCT-recovered shifts over $2\Delta t \in [4,48]$ hours drop below 2--3 pixels and most of the shifts are $\lesssim$1 pixels. Dashed lines in Figure \ref{F-opt_res_FLCT_DAVE4VM_2hr} show the optimization curves for 15-times rebinned 2-hour input (so that the $\sigma_{FLCT}$ and DAVE4VM window size are scaled to the original full resolution units). We can see that the optimal $\sigma_{FLCT}$ values are very similar between the full resolution and rebinned cases when the rebinned values are scaled to full resolution units, and consistent well within the $\sigma_{FLCT}$ grid spacing error of the rebinned case ($0.5\times 15 = 7.5$ pixels). This same result is recovered also for other cadences as illustrated by the second column of Table \ref{T-optimization_results}. Moreover, the rebinning introduces also a clear improvement in the $(C,S)$ metrics at the optimal $\sigma_{FLCT}$ values.

However, when it comes to the DAVE4VM window size, the optimization results are clearly not consistent between the full resolution and the 15-times rebinned cases, where the latter is strongly biased towards larger window sizes (Figure \ref{F-opt_res_FLCT_DAVE4VM_2hr}b and Table \ref{T-optimization_results}, last column). This is expected considering the fact that the smallest possible window size in the rebinned case is 3 rebinned pixels, and thus 45 full resolution pixels. Furthermore, since we consider only symmetric windows of odd window size in pixels, the next value in the grid is already 5 rebinned and 75 full resolution pixels. Since we employ the 5-point least-squares optimized derivatives in the spatial discretization of Eq. \ref{Eq-n_induc_eq} \citep{Jahne2004,Schuck2008,Liu2012}, and thus our estimates for the spatial derivatives at each pixel employ points over $5\times 5$ pixel area, it is also expected that the optimal results are $\geq$5 rebinned pixels ($\geq$75 pixels in full resolution). Despite this issue, we find that, similarly to the FLCT case, the metrics at optimal window sizes do improve after the rebinning (Figure \ref{F-opt_res_FLCT_DAVE4VM_2hr} and Table \ref{T-optimization_results}), as expected due to the markedly better discrete approximation of the induction equation.

\section{Supplementary figures}
	\label{S-appendix_supp_figs}
\begin{figure}[!h]  
    \centerline{\includegraphics[width=\textwidth, trim = 0.0cm 0.0cm 0cm 0.0cm]{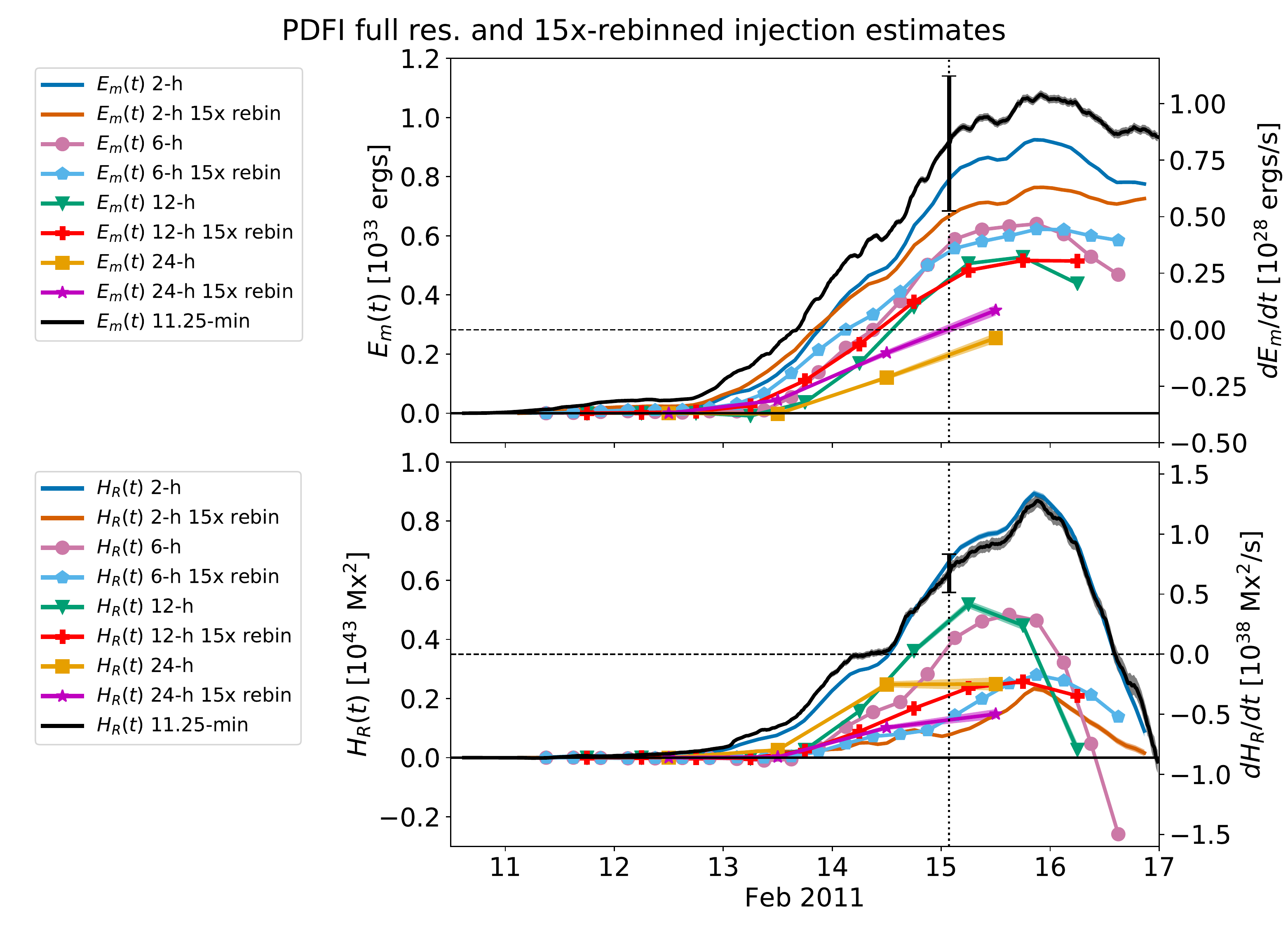}}  
   \caption{Energy $E_m(t)$ (upper panel) and helicity $H_R(t)$ (lower panel) injections for NOAA AR 11158 derived from PDFI electric field estimates for the lowest cadences ($\geq$2 h) using both full resolution and 15-times rebinned input data (solid curves), with error bars shown by the shaded regions surrounding the curves. The full-resolution 11.25-minute reference is also plotted with its combined noise- and method-related error bars at the time of the X-class flare (dotted black vertical line).} 
   \label{F-Em_HR_fluxes_PDFI_rebin}
\end{figure} 

\begin{figure}[!h]  
    \centerline{\includegraphics[width=\textwidth, trim = 0.0cm 0.0cm 0cm 0.0cm]{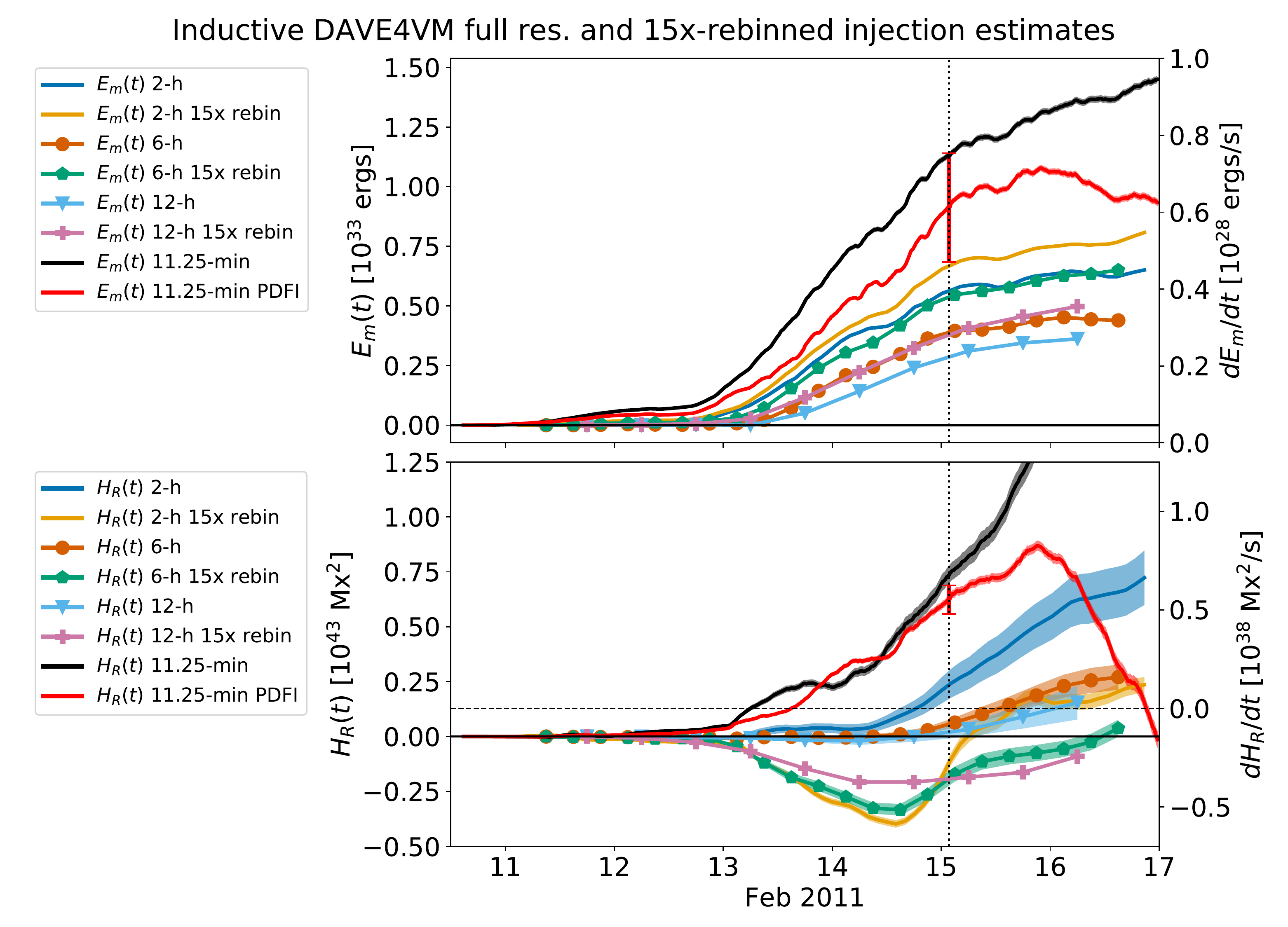}}  
   \caption{Energy $E_m(t)$ and helicity $H_R(t)$ injections for NOAA AR 11158 derived from inductive DAVE4VM electric field estimates for the lowest cadences ($\geq$2 h) using both full resolution and 15-times rebinned input data (solid curves), with error bars shown by the shaded regions surrounding the curves. The full-resolution 11.25-minute inductive DAVE4VM and PDFI estimates are also plotted with the combined noise- and method-related error bar for the latter at the time of the X-class flare (dotted black vertical line).} 
   \label{F-Em_HR_div_E_DAVE4VM_rebin}
\end{figure}

\FloatBarrier
\bibliographystyle{spr-mp-sola}
\bibliography{refs}  

\end{article} 

\end{document}